\documentclass[10pt,a4paper]{article}

\usepackage[utf8]{inputenc}
\usepackage{graphicx}
\usepackage{amssymb}
\usepackage{amsmath}
\usepackage{amsfonts}
\usepackage{amsthm}
\usepackage{wasysym}
\usepackage{color}
\usepackage{multirow}
\usepackage{url}
\usepackage[margin=1in]{geometry}
\usepackage[numbers,sort&compress]{natbib}
\usepackage{booktabs}

\title{Measuring the time-scale-dependent information flow between maternal and fetal heartbeats during the third trimester}

\author{Nicolas B. Garnier$^{1}$*, Sol Molinet$^{2}$, Marta Antonelli$^{3,4}$, \\Silvia Lobmaier$^{2}$, Martin Frasch$^{5,6}$*\\
\\
\small $^{1}$CNRS, ENS de Lyon, LPENSL, UMR5672, 69342, Lyon cedex 07, France\\
\small $^{2}$Department of Obstetrics and Gynecology, TUM University Hospital of Technical University of Munich, \\ \small TUM School of Medicine, Technical University of Munich, Germany\\
\small $^{3}$Technical University of Munich; Institute for Advanced Study, Garching, Germany\\
\small $^{4}$Instituto de Biología Celular y Neurociencia "Prof. Eduardo De Robertis", Facultad de Medicina, \\ \small Universidad de Buenos Aires, Argentina\\
\small $^{5}$Dept. of Obstetrics and Gynecology, University of Washington, Seattle, WA, USA\\
\small $^{6}$Institute on Human Development and Disability, University of Washington, Seattle, WA, USA\\
\\
\small *Correspondence: nicolas.garnier@ens-lyon.fr, mfrasch@uw.edu}

\begin{document}
\maketitle

\begin{abstract}
Prenatal maternal stress alters maternal-fetal heart rate coupling, as demonstrated by the Fetal Stress Index derived from bivariate phase-rectified signal averaging. Here, we extend this framework using information-theoretical measures to elucidate underlying mechanisms. In 120 third-trimester pregnancies (58 stressed, 62 control), we computed transfer entropy (TE), entropy rate (ER), and sample entropy (SE) under multiple conditioning paradigms, employing mixed linear models for repeated measures.
We identify dual coupling mechanisms at the short-term (0.5-2.5 s), but not long-term (2.5 - 5 s) time scales: (1) stress-invariant state-dependent synchronization, with maternal decelerations exerting approximately 60\% coupling strength on fetal heart rate complexity—a fundamental coordination conserved across demographics; and (2) stress-sensitive temporal information transfer (TE), showing exploratory associations with maternal cortisol that require replication. A robust sex-by-stress interaction emerged in TE from mixed models, with exploratory female-specific coupling patterns absent in males. Universal acceleration predominance was observed in both maternal and fetal heart rates, stronger in fetuses and independent of sex or stress.
We provide insight into the dependence of these findings on the sampling rate of the underlying data, identifying 4 Hz, commonly used for ultrasound-derived fetal heart rate recordings, as the necessary and sufficient sampling rate regime to capture the information flow. Information-theoretical analysis reveals that maternal-fetal coupling operates through complementary pathways with differential stress sensitivity, extending the Fetal Stress Index by elucidating causal foundations. Future studies should explore additional information-theoretical conditional approaches to resolve stress-specific and time-scale-specific differences in information flow.
%
\end{abstract}

\noindent\textbf{Keywords:} multiscale analysis ; information theory ; high-order statistics ; fetal heart rate ; causality

\section{Introduction}

Intermittent synchronization of maternal and fetal heartbeats has been established~\cite{Lobmaier:2020, VanLeeuwen:2003,VanLeeuwen:2009,DiPietro:2021}. Maternal-fetal heartbeat synchronization can serve as a biomarker of chronic stress during pregnancy, impacting maternal and fetal as well as postnatal well-being~\cite{Antonelli:2022}. Such findings are not only of scientific interest but also clinically actionable during pregnancy, helping increase awareness of chronic stress and inform behavioral and social interventions to reduce it. However, much remains unclear about this phenomenon. Specifically, its developmental physiology, the methodological and information-theoretical underpinnings of its computation remain active areas of research.

Here, we sought to apply the transfer entropy (TE) approach to elucidate the dynamics of the information flow between the mother and the fetus. Given the rising interest in wearables, which typically sample heart rate at low sampling rates to monitor wellness and clinical characteristics, we also sought to investigate the relationship between various TE-derived metrics and attributes of data acquisition and processing, with a focus on sampling rate and postnatal health outcomes. For this, we focused on the acceleration- and deceleration-dependent TE metrics.

We hope that our findings will contribute to the development or application of appropriate biosensor technologies to capture the dynamic properties of maternal-fetal heartbeat coupling effectively and to harness their potential to identify mother-fetus dyads at risk of suboptimal health outcomes, enabling earlier therapeutic intervention and steering health trajectories toward their optimal range.

\section{Methods}

\subsection{Cohort characteristics}

This prospective cohort study enrolled pregnant women receiving prenatal care between June 2016 and July 2019. Inclusion criteria were singleton pregnancy, gestational age 32--40 weeks at enrollment, and absence of known fetal anomalies. The study was approved by [IRB], and all participants provided written informed consent.

Maternal stress exposure was assessed using the Perceived Stress Scale (PSS; ~\cite{Cohen:1983}) and the Prenatal Distress Questionnaire (PDQ; ~\cite{Yali:1999}. Participants were classified as ``stressed'' (PSS $\geq$ 19) or ``control'' (PSS $<$ 19) based on established cutoffs for elevated perceived stress. Maternal hair cortisol concentration (pg/mg) was measured as a physiological marker of chronic stress exposure.

The initial findings of the underlying FELICITy study, using the bPRSA approach and the SAVEr algorithm for fetal/maternal ECG and R peak extraction, have been reported elsewhere~\cite{Lobmaier:2020, Sarkar:2022, Li:2017}.

Ethics approval was obtained from the Committee of Ethical Principles for Medical
Research at the TUM (registration number 151/16S; ClinicalTrials.gov registration number NCT03389178). All methods were performed in accordance with the relevant guidelines and regulations. We obtained informed consent for study participation from each subject. The complete experimental design is reported in ~\cite{Lobmaier:2020}.

\begin{figure}[htb]
\begin{center}
\includegraphics[width=.9\linewidth]{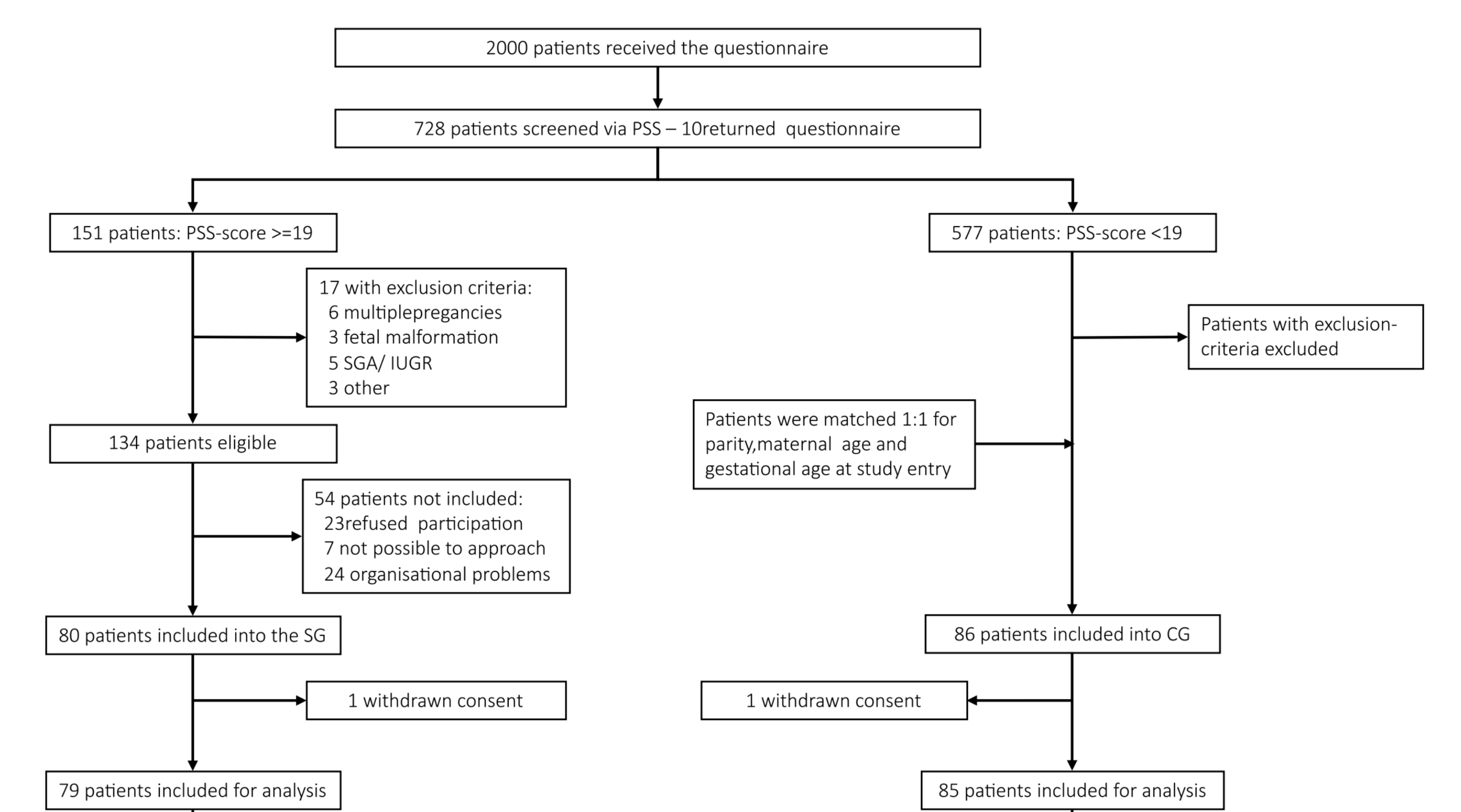}
\end{center}
\caption{Recruitment flow chart for the FELICITy dataset.}
\label{fig:cohort}
\end{figure}

Stressed mothers were matched 1:1 with controls on parity, maternal age, and gestational age at study entry. Recruited subjects were between 18 and 45 years of age and were in their third trimester. The subjects were selected from a cohort of pregnant women followed in the Department of Obstetrics and Gynecology at "Klinikum rechts der Isar" of the Technical University of Munich (TUM). This is a tertiary
center of Perinatology located in Munich, Germany, which serves 2000 mothers/newborns per year. Fig.~\ref{fig:cohort} presents the recruitment flowchart for this dataset and the data used in this study.

Four exclusion criteria were applied, namely (a) serious placental alterations defined as fetal growth restriction according to Gordijn et al.~\cite{Gordijn:2016};
(b) fetal malformations; (c) maternal severe illness during pregnancy; (d) maternal drug or alcohol abuse.

The Cohen Perceived Stress Scale questionnaire was administered to gauge chronic non-specific stress exposure (PSS-10)~\cite{Cohen:1983}. PSS-10 $\geq$ 19 categorized subjects as stressed, as established~\cite{Lobmaier:2020}. We applied inclusion and exclusion criteria following the return of the questionnaires. When a subject was categorized as stressed, the next screened participant, matched on gestational age at recording and with a PSS-10 score < 19, was entered into the study as a control. In addition to PSS-10, the participants received the German Version of the "Prenatal Distress Questionnaire" (PDQ), which contains 12 questions on pregnancy-related fears and worries regarding changes in body weight and other pregnancy-related issues, the child's health, delivery, and the impact of pregnancy on the woman's relationship.

\subsection{Fetal and Maternal Heart Rate Acquisition}
\label{sec:aquisition}

Simultaneous fetal and maternal ECG recordings were obtained as transabdominal ECG (aECG). We standardized the clinical setting as much as possible across all study participants. We performed the recordings on all women in the supine half-left recumbent position, usually at the same time of day (afternoon). The aECG was recorded at a 900 Hz sampling rate for at least 40 min at 2.5 weeks after initial screening. The AN24 (GE HC/Monica Health Care, Nottingham, UK) was used. We calculated the signal quality index (SQI)~\cite{Li:2017} for aECG in 1-s windows and subsequently discarded segments with an SQI below 0.5. Using the fetal and maternal ECG deconvolution algorithm SAVER~\cite{Li:2017}, we extracted fetal ECG (fECG) and maternal ECG (mECG) at a 1000 Hz sampling rate.

We detected fetal and maternal R-peaks separately from the corresponding fECG and mECG signals. The fetal- and maternal RR interval time-series were subsequently derived from the fetal and maternal R-peaks. We then calculated the mean fetal heart rate (fHR) and mean maternal heart rate (mHR) values.

Upon delivery of the baby, we recorded the clinical data, including birth weight, length, and head circumference, pH, and Apgar score.

Next, we removed 44 mother-foetus dyads from the cohort because for these dyads the fHR was of the order of the mHR (i.e., below 100 bpm) for a significant fraction of the experiment.

HR data were computed from R-R intervals extracted from the ECG signal, with R peaks identified by the SAVEr algorithm at 1000 Hz resolution. At this stage, there was a noise of the order of 0.5 bpm in the HR data due to the $f_{\rm ECG}=1000$ Hz sampling rate of ECG.

\subsection{Low-pass filtering}
\label{sec:filtering}

\begin{figure}[htb]
\begin{center}
\includegraphics[width=.75\linewidth]{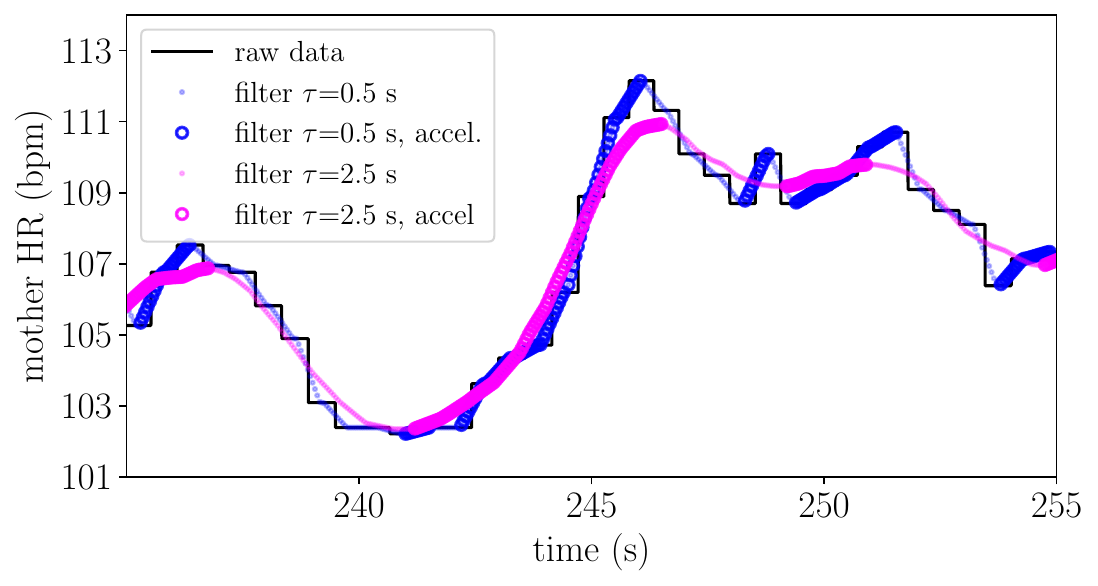}
\end{center}
\caption{Example of mother HR (mHR) data.
The raw mHR sampled at 1kHz is a stepwise function of time (black). We low-pass filter the raw mHR signal using a low-pass cutoff frequency $1/\tau$, where the time-scale $\tau$ represents the scale at which we will further analyze the information contents of the heart rates. Low-pass-filtered signal are then downsampled at $f_s$=20Hz.
Two examples are represented: one for $\tau$=0.5s (blue) and one for $\tau$=2s (magenta).
Larger circles indicate times when the filtered mHR signal is increasing, which we define as {\em accelerations}. Conversely, little dots correspond to {\em decelerations}.}
\label{fig:filtering}
\end{figure}

The raw heart rate $X$ signal is derived from the successive time positions of the R-peaks of the ECG signal, given by {SAVEr} with a resolution $f_{\rm ECG}$, as follows.
This raw HR signal is a step-wise function of time sampled at $f_{\rm ECG}$: its value is constant between two consecutive R peaks. Indeed, as time passes, its value changes when, and only when, a new R peak occurs: it is then possible to compute the size of the RR-interval that just ended when the new R peak occurs. A new constant value of the raw heart rate can then be deduced, which is a continuous value for all RR-intervals.
To have a continuously evolving HR signal sampled at a given frequency $f_s$ ---~possibly much lower than the typical frequency associated with RR-intervals~---, we first low-pass filter the raw HR signal by using a local averaging over a time interval corresponding exactly to the timescale $\tau$ we are willing to study.
\begin{align}
X_\tau(t) &= \frac{1}{\tau}\int_{t'=t-\tau}^t X(t') dt' \label{eq:avg:formal}\\
&= \frac{1}{\tau f_{\rm ECG}} \sum_{k=k_{t-\tau}+1}^{k_t} X_k \label{eq:avg:practical} \,,
\end{align}
where we have noted $k_t = t f_{\rm ECG}$ the index of the point at time $t$ in a signal sampled at $f_{\rm ECG}$. The first eq.(\ref{eq:avg:formal}) is formal and relates to a time-continuous signal while the second eq.(\ref{eq:avg:practical}) is the one used in practice.
This procedure is depicted in Fig.~\ref{fig:filtering}.

Filtering removes noise and information at frequencies higher than the cutoff frequencies $1/\tau$ which correspond to the timescale $\tau$ we are studying. For example, the SAVER algorithm (section~\ref{sec:aquisition}) is upsampling to 1000 Hz the aECG data from AN24 recorded at 900 Hz; this interpolation introduces some higher frequency noise, which is removed by the filtering.
In the following, we will vary $\tau$ in the range [0.5s-20s] to explore the distribution of information in the heart rates. The lowest value $\tau=0.5$s in this range corresponds to the typical time interval between two consecutive R-peaks: examining time-scales lower than 0.5s is thus illusory, as there is no information in the raw HR signal at these scales.
The larger value $\tau=20$s is sufficient to retain all the interesting behavior of the interactions we are searching for.

After filtering, the data $X_\tau$ is down-sampled at a fixed sampling frequency $f_s$ which we set to 20Hz, unless noted otherwise in section ~\ref{sec:sampling_effect} where we study the effect of the sampling frequency.

\subsection{HR Decelerations and Accelerations}

\subsubsection{Definition of HR decelerations and accelerations}

We define HR accelerations and decelerations~\cite{Piskorski:2011} at a given time-scale $\tau$ as follows.
For a given HR signal $X$, we first low-pass filter this signal as described above using the time-scale $\tau$, and then examine the sign of the time-derivative of the filtered signal.
We define accelerations ${\cal A}_\tau$ as the set of times (epochs) where the time-derivative of the filtered signal $X_\tau$ is positive:
\begin{equation}
t \in {\cal A}_\tau \Leftrightarrow \frac{d X_\tau(t)}{dt} > \mu, \qquad {\rm with} \quad \mu=0
\label{def:accel}
\end{equation}
Respectively, we define decelerations ${\cal D}_\tau$ as the set of times where the time-derivative is negative:
\begin{equation}
t \in {\cal D}_\tau \Leftrightarrow \frac{d X_\tau(t)}{dt} < -\mu, \qquad {\rm with} \quad \mu=0
\label{def:decel}
\end{equation}

Because the value of the time-derivative depends on the time-scale of the filter, so does the partitioning of epochs in accelerations and decelerations, as can be seen in Fig.~\ref{fig:filtering}. At a given timescale $\tau$, accelerations, resp. decelerations occur in sets of consecutive times (epochs), the duration of which is typically larger than $\tau$. As a consequence, for smaller time-scales $\tau$ we expect a larger number of distinct ---~{\em i.e.}, well-separated in time~--- accelerations, resp. decelerations, and for larger time-scales we expect a smaller number of distinct accelerations, resp. decelerations, while each is expected to be "longer" in that it should contain more points.

Our method can be tuned by requiring that the absolute value of the time derivative exceed a finite threshold $\mu\neq0$. Still, our explorations showed that increasing the threshold severely reduces the number of points in acceleration and deceleration epochs. This is expected, as increasing $\tau$ smoothes the HR signal more, thus reducing its dynamics: the standard deviation of $X_\tau$, and hence of its time-derivative, is typically $\sqrt{\tau}$ times smaller than the standard deviation of $X$. In this article, we chose a threshold $\mu=0$ to ensure sufficient points for accelerations and decelerations.

\subsubsection{Quantification of acceleration/deceleration events}

Heart rate accelerations and decelerations were identified for both maternal and fetal HR. For each recording, we computed the total number of acceleration events (N\_accel) and deceleration events (N\_decel), allowing calculation of event fractions:

\begin{itemize}
    \item Acceleration fraction = N\_accel / (N\_accel + N\_decel)
    \item Deceleration fraction = N\_decel / (N\_accel + N\_decel)
\end{itemize}
These were computed separately for maternal HR conditioning and fetal HR conditioning.

\subsection{Information, Complexity and Information Flow between Maternal and Fetal HR}

\subsubsection{Conditioning Framework}

We computed univariate entropy metrics (hmax and hmean) at three levels of analysis to characterize both univariate signal properties and bivariate maternal-fetal coupling:

\begin{enumerate}
    \item Univariate baseline: \textbf{Entropy rate (ER)/Sample Entropy (SE)} computed on the full fetal or maternal HR time series without conditioning (no\_conditioning), capturing baseline complexity of each signal independently.
    \item Cross-conditioned bivariate: ER/SE of one signal (e.g., fetal HR) computed specifically during events detected in the other signal (e.g., maternal accelerations or decelerations). This framework inherently captures maternal-fetal coupling: if fetal entropy differs when conditioned on maternal events versus no conditioning, this reveals that mHR state modulates fHR complexity - a signature of physiological interdependence.
    \item Self-conditioned: ER/SE of a signal during its own detected events (e.g., fetal HR during fetal accelerations), capturing state-dependent complexity within the same signal.
\end{enumerate}

\subsubsection{Entropy metrics: overview}

Three categories of entropy-based features were computed from the mHR and fHR time series:

\textbf{ER:} Entropy rate quantifies the complexity or unpredictability of a time series, representing the rate at which new information is generated. A higher entropy rate indicates greater irregularity in HR dynamics. ER was computed separately for fetal and maternal HR signals.

\textbf{SE:} Sample entropy measures the regularity of a time series by quantifying the conditional probability that sequences similar for $m$ points remain similar at $m+1$ points~\cite{Richman:2000b,Lake:2002}. Lower SE indicates more regular, predictable patterns. SE was computed with embedding dimension $m=1$ and tolerance $r=0.2 \sigma$, where $\sigma$ is the standard deviation of the signal under study. SE was computed separately for fetal and maternal HR signals.

\textbf{Transfer Entropy (TE):} Transfer entropy quantifies the directed information flow between two time series, measuring the extent to which knowledge of one signal reduces uncertainty about the future of another. TE was computed bidirectionally between maternal and fetal HR, with conditioning on either fetal HR (TE\textsubscript{fHR}) or maternal HR (TE\textsubscript{mHR}).

All features were computed using our own toolbox~\cite{Garnier:2025}.
For each feature $A$, both maximum $A^{\rm max}$ and mean $A^{\rm AUC}$ values were extracted using the time-scale interval $\tau \in [0.5, 2.5]$ seconds. The maximum value is computed as the maximum of the feature $A$ in the interval and is labeled $A^{\rm max}$ or refered to as max $A$. The mean value is computed as the average of the feature $A$ over the interval which, up to a factor 2$s$ is exactly the area under the $A(\tau)$ curve (see Fig.~\ref{fig:complexities} for such curves for ER and Fig.~\ref{fig:varying_fs:average} for TE); we label it $A^{\rm AUC}$ and refer to it as mean $A$.

Following our conditioning framework, these were computed under five conditioning paradigms: (1) full recording, (2) fetal HR acceleration epochs, (3) fetal HR deceleration epochs, (4) maternal HR acceleration epochs, and (5) maternal HR deceleration epochs. This yielded

\begin{itemize}
    \item 20 ER features: max/mean $\times$ fetus/mother $\times$ \{full, fHR\_accel, fHR\_decel, mHR\_accel, mHR\_decel\}
    \item 20 SE features: Same structure as ER
    \item 12 TE features: max/mean $\times$ \{fHR/mHR conditioning\} $\times$ \{all, accel, decel\}
    \item Total: 52 entropy-based features
\end{itemize}

\subsubsection{Entropy rate}

To get some insight into the information content of the conditioned signals, we measured their entropy rate~\cite{Spilka:2014,Granero:2017,Granero:2019} over the time-scale $\tau$:
\begin{align}
h(\tau) &=H(X_\tau(t),X_\tau({t-\tau}))-H(X_\tau({t-\tau})) \\
&= H^{(2)}(X_\tau)-H(X_\tau)
\end{align}
where $H(X_\tau)$ is the Shannon entropy of the filtered signal $X_\tau$ and $H^{(2)}(X_\tau)$ is the Shannon entropy of the bi-variate filtered signal $(X_\tau(t), X_\tau(t-\tau))$, obtained by time-embedding the filtered signal $X_\tau$ over the time-scale $\tau$.
Unless noted otherwise, this study uses filtered data sampled at $f_s=$20Hz.

We compute the entropy rate for a set of time-scales $\tau$ ranging from $1/fs$=0.05 seconds up to 20 seconds.
Because we are interested in time-scale range [0.5; 2.5]s, we compute the maximal value $h^{\rm max}$ of the entropy rate in that range, as well as its mean value $h^{\rm AUC}$ in the same range, which up to a factor 2s represents the area under the curve (AUC) of $h(\tau)$ in the range  [0.5; 2.5]s.

\subsubsection{Transfer Entropy}

To explore the respective influences of mother and foetal heart rates, we compute the transfer entropy between the filtered heart rates of the mother and her foetus.
Transfer entropy was introduced by Schreiber~\cite{Schreiber:2000} and has since gained widespread popularity for studying information exchange between two signals across a wide range of systems.

For two signals $X(t)$ and $Y(t)$ and a positive time lag $\tau'$, transfer entropy ${\rm TE}^{(\tau')}(X(t)\rightarrow Y(t))$ expresses the amount of shared information between $X(t)$ and $Y(t+\tau')$ that is not contained in $Y(t)$. In other words, it quantifies how much information in the future of $Y$ at time $t+\tau'$ is also present in the present of $X$ at time $t$, while not being present in $Y$ itself at time $t$.
It is thus interpreted as the amount of information that flows from $X$ to $Y$ on the time-scale $\tau'$.

Noting $M_\tau$ and $F_\tau$ the maternal and the fetal heart rates filtered at the time-scale $\tau$ according to the protocol described in section~\ref{sec:filtering}, we naturally set the time-lag $\tau'=\tau$ to define
\begin{align}
{\rm TE}_{m \rightarrow f}(\tau) &= {\rm TE}^{(\tau)}(M_\tau(t) \rightarrow F_\tau(t)) \\
{\rm TE}_{f \rightarrow m}(\tau) &= {\rm TE}^{(\tau)}(F_\tau(t) \rightarrow M_\tau(t))
\end{align}

We also define the "net" TE from mother to fetus as the difference:
\begin{equation}
{\rm TE}(\tau)={\rm TE}_{m \rightarrow f}(\tau)-{\rm TE}_{f \rightarrow m}(\tau)\,,
\end{equation}
which is positive when more information flows from the mother to her foetus than information flows from the foetus to the mother, and negative otherwise.

The TE is measured using a nearest neighbors estimator with $k=5$ neighbors~\cite{Kraskov2004} and using 4000 points. We estimate the bias of this estimator by computing the TE with surrogate data, as follows. For each set of signals $(X,Y)$, we construct 10 sets of surrogate signals $(X_s, Y_s)$ by shuffling the original signals, which destroys the complete time dependencies between $X$ and $Y$. We then measure TE$(X_s \rightarrow Y)$ and TE$(Y_s \rightarrow X)$. The values we obtain are very small (of the order $10^{-3}$). We average these values over the set of 10 surrogates and subtract this estimate of the bias from the TE$(X\rightarrow Y)$ and TE$(Y\rightarrow X)$, respectively.

Because a dataset contains much more than 4000 points in time, we compute the TE on 10 different subsets of 4000 points, randomly picked in the dataset. This allows us to estimate the standard deviation of the TE estimator, which we use to represent error bars in the figures. For a given subject, these error bars are small.

We compute the TE for a set of time lags $\tau$ ranging from $1/f_s$ up to 20$f_s$, with a step $1/f_s$. Because we are interested in time-scale range $[0.5; 2.5]$s, we compute the maximal value TE$^{\rm max}$ of the TE in that range, as well as the mean value TE$^{\rm AUC}$ of the TE within that range, which up to a factor 2s represents the area under the curve (AUC) of TE$(\tau)$ in that range.

\subsection{Neurodevelopmental Outcome Assessment}

Infant neurodevelopmental outcomes were assessed at 24 months using the Bayley Scales of Infant and Toddler Development, Fourth Edition (Bayley-4). Composite scores were obtained for cognitive (COG), language (LANG), and motor (MOTOR) domains\cite{Molinet:2026}. Subscale scores for language (receptive, expressive) and motor (fine, gross) skills were also analyzed.

\subsection{Statistical Analysis}

\subsubsection{Correlation Analysis (TE/ER/SE vs Clinical Outcomes)}

\textbf{Normality Assessment}: Distributional properties of all variables were assessed using the Shapiro-Wilk test ($\alpha$ = 0.05). Correlation method selection was data-driven: Pearson product-moment correlation was applied when both variables satisfied normality assumptions; Spearman rank correlation was used otherwise.

\textbf{Univariate Correlations}: Bivariate correlations were computed between each entropy feature (TE, ER, SE) and outcome variable (cortisol, Bayley scores, PSS, PDQ). This yielded 144 independent correlation tests across feature-outcome combinations. Given the exploratory nature of this analysis, uncorrected p-values are reported with nominal significance threshold of p < 0.05 (two-tailed). However, we also report false discovery rate (FDR) corrected q-values using the Benjamini-Hochberg procedure to assess robustness to multiple comparison correction. At the expected false positive rate of 5\%, approximately 7.2 spurious significant findings would be anticipated by chance alone. All findings should be interpreted as hypothesis-generating rather than confirmatory and require independent replication.

\textbf{Stratified Analyses}: To examine potential moderating effects, correlations were computed separately for: (1) stressed versus control groups based on PSS classification, and (2) male versus female fetuses.

\subsubsection{Mixed Linear Model Analysis (Repeated Measures)}

To properly account for repeated measures within subjects and enable interaction testing, we employed mixed linear models (MLMs) with restricted maximum likelihood (REML) estimation for three separate analyses:

\medskip

\textbf{Model 1 - Acceleration/Deceleration Ratios}:
\begin{align*}
{\tt Fraction} \sim {\tt Sex} \times {\tt Stress} \times {\tt HR\_Source} \times {\tt Event\_Type} + {\tt (1|Patient\_ID)}
\end{align*}

Data structure: 480 observations (120 patients $\times$ 4 measurements each: mHR\_accel, mHR\_decel, fHR\_accel, fHR\_decel). Random intercept accounts for patient-level correlation.

\medskip

\textbf{Model 2 - Entropy Rate with Conditioning}:
\begin{align*}
{\tt Value} \sim {\tt Sex} \times {\tt Stress} \times {\tt Metric} \times {\tt HR\_Source} \times {\tt Conditioning} + {\tt (1|Patient\_ID)}
\end{align*}

Data structure: 1,262 observations (average 10.5 per patient). Conditioning levels included: none (univariate baseline), mother\_accel, mother\_decel, fetus\_accel, and fetus\_decel (cross-conditioned bivariate measures). Selected 2-way interactions included based on theoretical relevance.

\medskip

\textbf{Model 3 - Sample Entropy with Conditioning (Methodological Assessment):}
\begin{align*}
{\tt Value} \sim {\tt Sex} + {\tt Stress} & + {\tt HR\_Source} + {\tt Conditioning} \\
 & + {\tt HR\_Source:Conditioning} + {\tt (1|Patient\_ID)}
\end{align*}

\medskip

Sample entropy MLM analysis was attempted using the same conditioning framework as the entropy rate. However, due to the sample entropy's requirement for minimum data points, 87-100\% of the values were zero in conditioned windows (accelerations/decelerations lasting 2-10 seconds). This necessitated a simplified model that included only conditioning type with \(\geq 5\%\) non-zero values: none (baseline), mother\_accel (13.3\% non-zero), and mother\_decel (5.8\% non-zero). Data structure: 286 observations (2.4 per patient, vs.\ 10.5 for entropy rate).

\textbf{Rationale for MLM}: Each patient contributes multiple measurements, creating within-subject correlation. Independent t-tests would treat these as independent observations, leading to inflated effective sample size, underestimated standard errors, and artificially low p-values (pseudoreplication). Random intercepts for Patient\_ID provide valid statistical inference for hierarchical data.

All models included two-way interactions between fixed effects. Statistical significance was assessed at \(\alpha = 0.05\). Analyses were conducted in Python 3.9 using statsmodels 0.14.4.

\subsubsection{Multivariate Modeling}

To assess whether entropy features jointly predict neurodevelopmental outcomes and whether maternal stress moderates these relationships, we applied multiple regularization and dimensionality reduction approaches to handle the high-dimensional feature space (52 predictors: 20 ER + 20 SE + 12 TE, plus stress indicator).

\medskip

\textbf{Methods Applied:}

\begin{enumerate}
    \item \textbf{Elastic Net Regression}: Combined L1 (Lasso) and L2 (Ridge) penalties to handle multicollinearity and perform automatic feature selection. The mixing parameter $\alpha$ was optimized via 5-fold cross-validation.
    \item \textbf{PCA + Ridge Regression}: Principal component analysis followed by Ridge regression to address multicollinearity through orthogonal transformation of features. Components explaining 95\% cumulative variance were retained.
    \item \textbf{Partial Least Squares (PLS) Regression}: Designed specifically for high-dimensional, multicollinear predictors. Identifies latent components that maximize covariance between features and outcomes.
    \item \textbf{Random Forest Regression}: Non-parametric ensemble method capturing non-linear relationships and interactions without assumptions about feature distributions.
    \item \textbf{Parsimonious Forward Selection}: Ordinary least squares regression with forward feature selection limited to a maximum of 3 features, appropriate for small sample sizes.
\end{enumerate}

\medskip

\textbf{Pre-processing}: All features were standardized (z-scored) before modeling. Cross-validation employed 5-fold CV where sample size permitted (reduced for outcomes with $n<20$).

\medskip

\textbf{Multicollinearity Assessment}: Variance Inflation Factor (VIF) was calculated using statsmodels to quantify multicollinearity among predictors. VIF > 10 indicates severe multicollinearity requiring regularization approaches.

\medskip

\textbf{Sample Size Considerations}: The ratio of sample size to number of predictors (n/k ratio) should ideally exceed 10-20 for reliable multivariate inference. With neurodevelopmental outcomes available for $n = 30-66$ participants and 53 total predictors (52 features + stress), the $n/k$ ratios ranged from 0.6 to 1.2, indicating severe underpowering for standard regression approaches.

\section{Results}

\subsection{Study sample characteristics}

After signal pre-processing steps, a total of 120 mother-fetus dyads with complete entropy feature data were included from the initial 164. The cohort was balanced for maternal stress exposure: 62 participants (51.7\%) were classified as controls and 58 (48.3\%) as stressed. The sample included 49 male (40.8\%) and 71 female (59.2\%) fetuses.

Neurodevelopmental follow-up data were available for a subset of participants: cognitive composite scores ($n$ = 66, 55.0\%), language composite scores ($n$ = 63, 52.5\%), and motor composite scores ($n$ = 65, 54.2\%). Hair cortisol data were available for 88--90 participants depending on the analysis.

\subsection{Decelerations and accelerations: dependence on stress and time scale}

An important feature of our approach is that decelerations and accelerations are defined using filtered signals and varying the time scale. This is in strong contrast to a more classical definition using the raw heart rate signal ---~which is piece-wise constant~--- which would require using a non-zero threshold $\mu$.

First, we explore the distribution of decelerations and accelerations as a function of conditioning approach (on mother or on fetus), exposure, and sex. Next, we will explore the information-theoretical properties.

\begin{figure}[htb]
\begin{center}
\includegraphics[width=.9\linewidth]{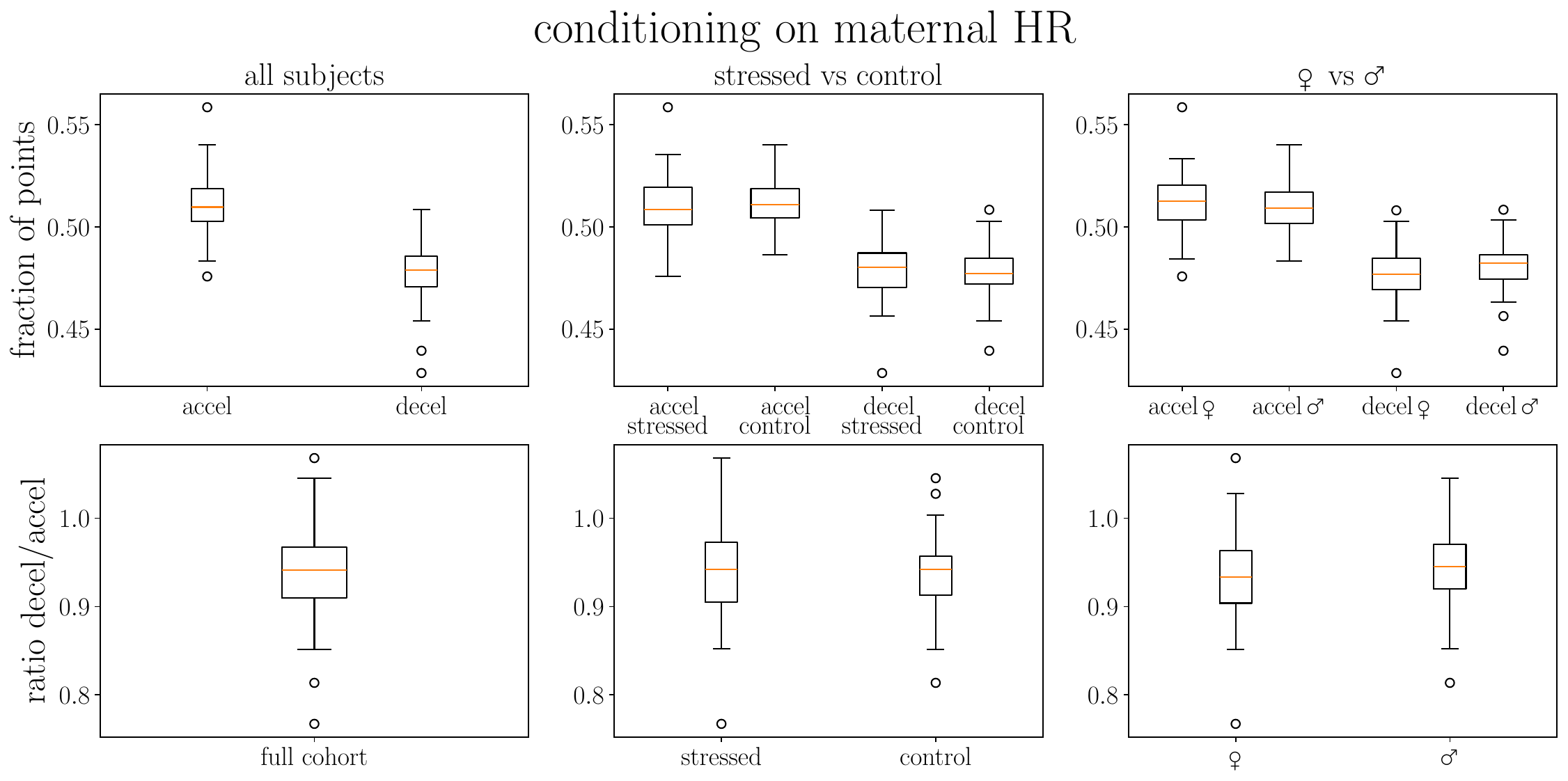} \\
\includegraphics[width=.9\linewidth]{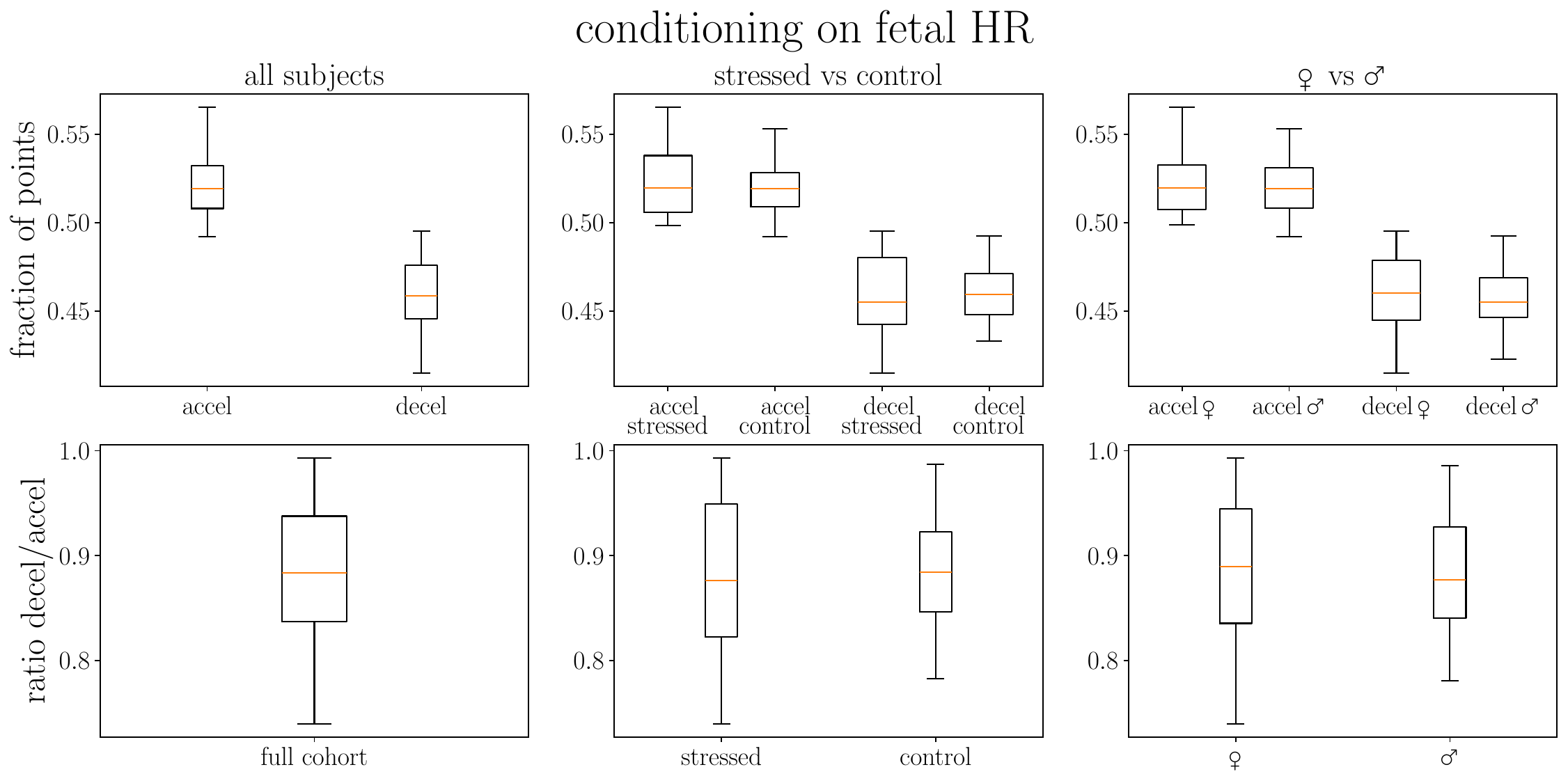}
\end{center}
\caption{Statistics of decelerations/accelerations ratios computed at timescale $\tau=2.5$s.
The upper and lower parts correspond to the mother's HR and the fetus's HR, respectively.
In each case, the first line presents the fraction of time points that are decelerations or accelerations. In contrast, the second line presents the ratio of decelerations to accelerations.
}
\label{fig:accel_decel_stats}
\end{figure}

We present in Fig.~\ref{fig:accel_decel_stats} and Fig.~\ref{fig:accel_decel_summary} the distribution of the ratio of decelerations to accelerations in the cohort.
The ratios of decelerations/accelerations are always around one. The ratio below 1 indicates the dominance of accelerations. Interestingly, we observe slightly fewer decelerations than accelerations, especially for fetal heart rates, consistent with previous literature~\cite{Pawlowski:2025}.

\subsubsection{Overall Patterns }

Mixed linear model analysis revealed a significant main effect of event type ($\beta$ = -0.061, SE = 0.0028, p < 0.001) and HR\_Source $\times$ Event\_Type interaction ($\beta$ = 0.028, SE = 0.0028, p < 0.001).

Accelerations were significantly more common than decelerations across all conditions. Asymmetry varied by heart rate source (Figure~\ref{fig:accel_decel_stats}):
    \begin{itemize}
        \item Fetal HR: 52.1\% accel vs 46.0\% decel (6.1\% difference)
        \item Maternal HR: 51.1\% accel vs 47.8\% decel (3.3\% difference)
    \end{itemize}

HR\_Source showed a main effect ($\beta$ = -0.011, p < 0.001), i.e., mHR shows a different baseline pattern.

\subsubsection{Group Comparisons }

\begin{figure}[htb]
\begin{center}
\includegraphics[width=.9\linewidth]{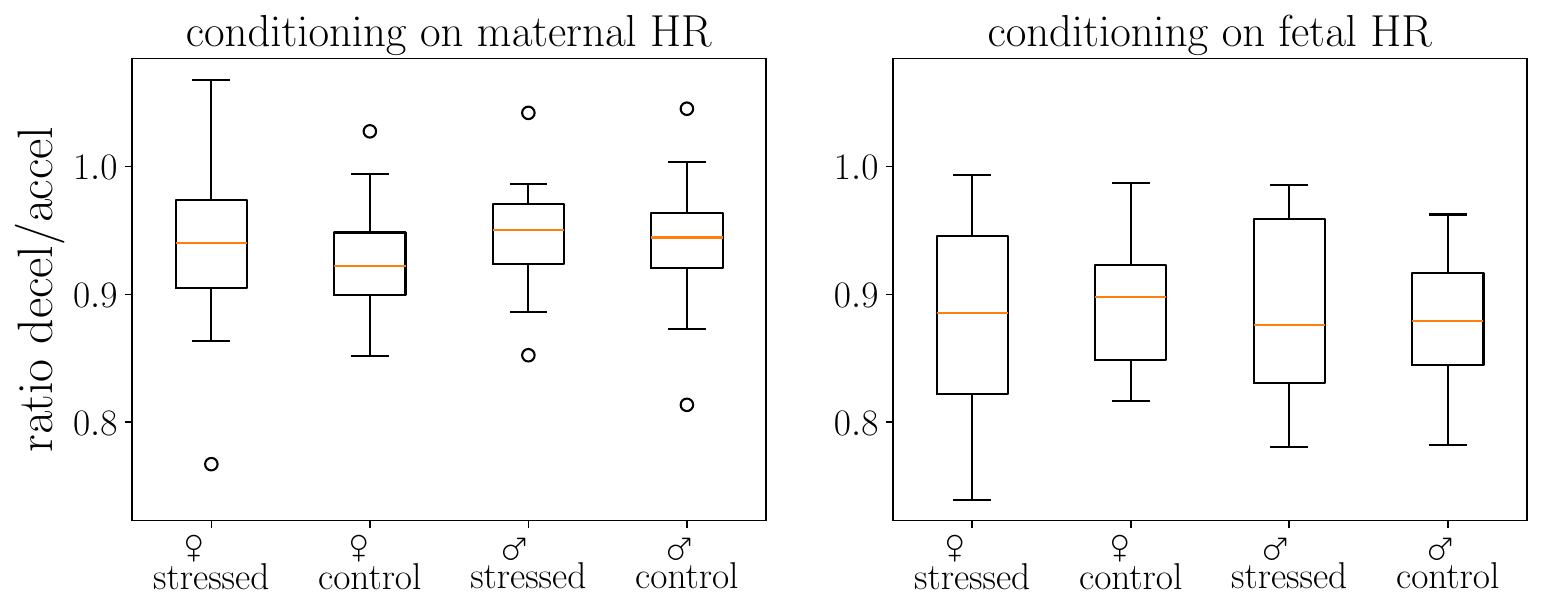}
\end{center}
\caption{Group dependencies of statistics of decelerations/accelerations ratios computed at timescale $\tau=2.5$s, for female (\female) and male (\male) fetuses, in both the stressed and control group.
}
\label{fig:accel_decel_summary}
\end{figure}

We observed no significant demographic effects (Figure~~\ref{fig:accel_decel_summary}):

\begin{itemize}
    \item Sex (male): $\beta$ = -0.0026, p = 0.361
    \item Stress (stressed): $\beta$ = +0.0007, p = 0.802
    \item Sex $\times$ Stress interaction: $\beta$ = -0.0001, p = 0.985
\end{itemize}

Acceleration/deceleration asymmetry represents a universal biological phenomenon, consistent across demographic groups in this third-trimester cohort. Full statistical results are presented in \ref{tab:mlm_acc_dec_complete}.

\subsection{Univariate information measures of fHR and mHR: Entropy Rate and Sample Entropy}

\subsubsection{Identification of time scales}

\begin{figure}[htb]
\begin{center}
\includegraphics[width=.9\linewidth]{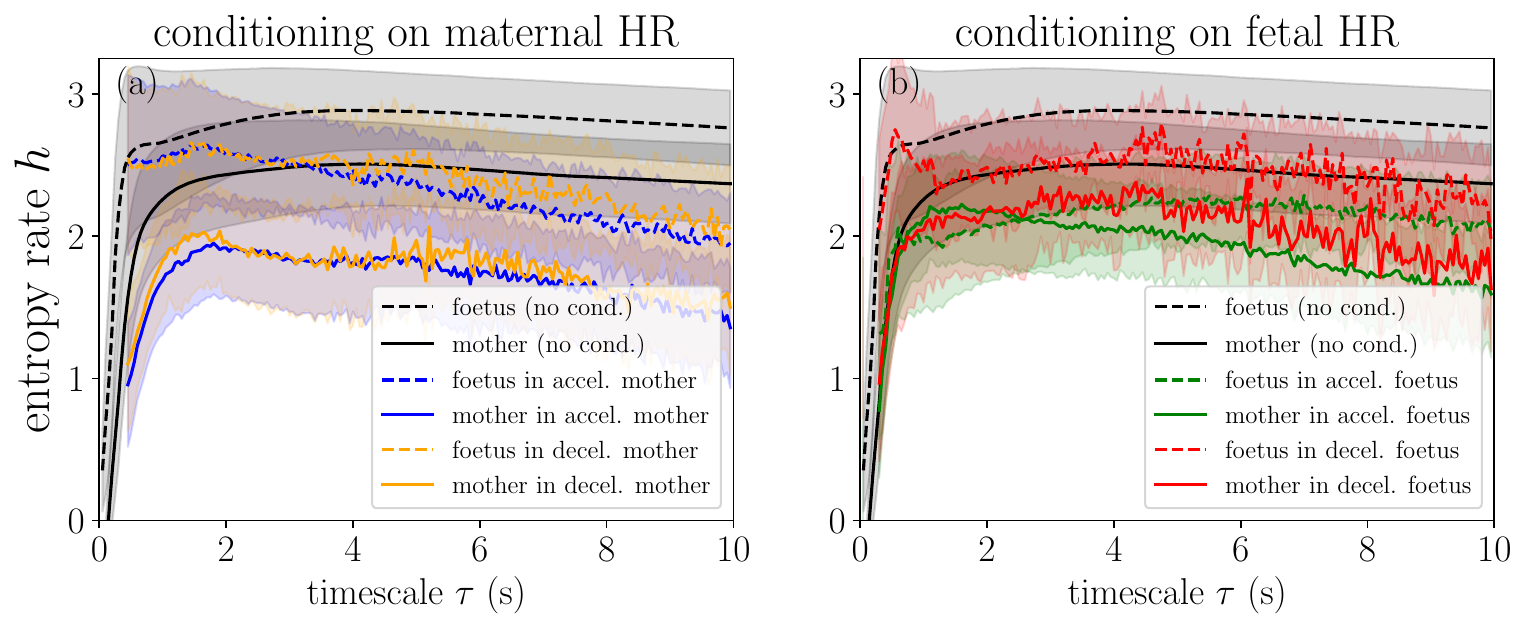}
\end{center}
\caption{Entropy rate $h$ as a measure of information or complexity in the conditioned signals. (a) results when conditioning is performed on the maternal HR signal, and (b) when it is performed on the fetal HR signal. Dotted lines pertain to fetal HR and continuous lines to maternal HR. Blue (resp. orange) curves are obtained by conditioning on mHR accelerations (resp. mHR decelerations). Green (resp. red) curves are obtained by conditioning on fHR accelerations (resp. fHR decelerations). For reference, we have plotted the entropy rate measured in fHR (dotted black line) and mHR (continuous black line) without any conditioning. For each curve, a shaded area represents the dispersion of the values over the cohort.
}
\label{fig:complexities}
\end{figure}

We present in Fig.~\ref{fig:complexities} these quantities averaged over the cohort. For each quantity, we estimate for each timescale a dispersion interval as the standard deviation of the quantity over the cohort (shaded areas in Fig.~\ref{fig:complexities}); this dispersion interval is typically of $\pm0.5$, in entropy rate units (nats).
We observe that the entropy rate is roughly constant across all time-scales, except for time-scales below 0.5-1s. This is expected, as there is no information in the HR signal between consecutive RR peaks, which are typically separated by 0.5s for fHR and 0.8s for mHR.

There are no prominent TE maxima. We can nevertheless deduce that the exchange occurs on a timescale shorter than 5s.

When no conditioning is used, the entropy rate is always greater than that of any counterpart with conditioning; this holds for both fHR and mHR signals.
We observe that whether with or without conditioning, the fHR always has a higher entropy rate than the mHR, except when conditioning using the fHR accelerations. In this latter case only, fHR appears to have a lower entropy rate than mHR in the band [0.5-2.5] s.

\subsubsection{Stress and sex effects}

\begin{figure}[htb]
\begin{center}
\includegraphics[width=.9\linewidth]{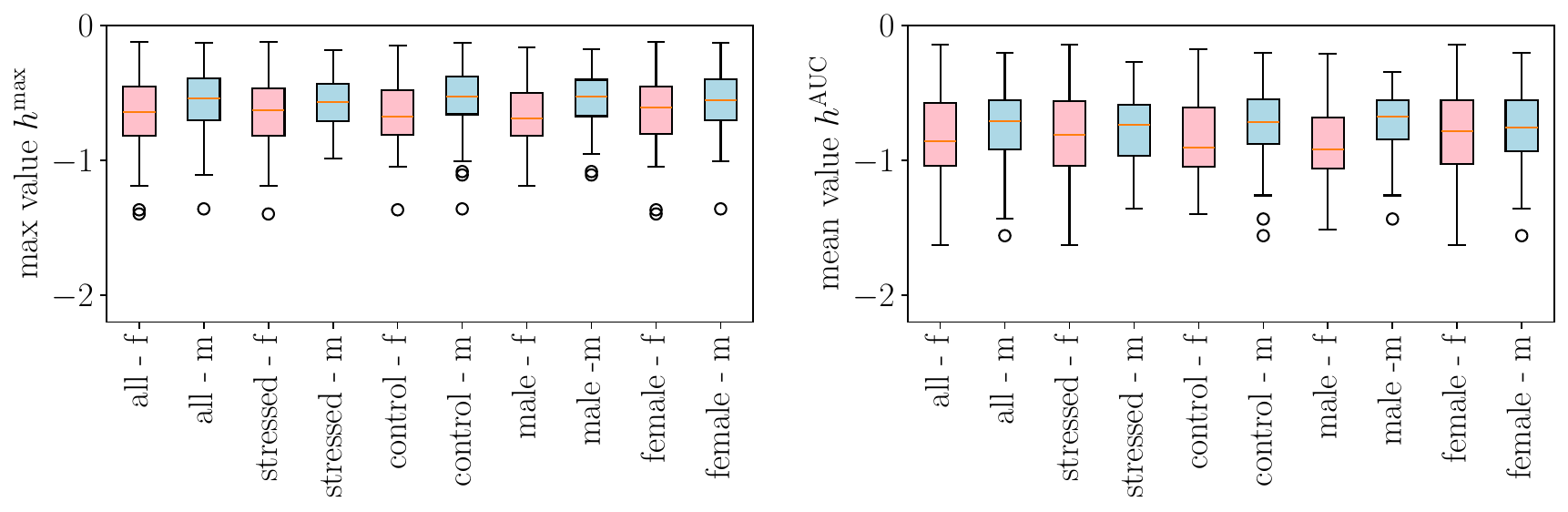}
\end{center}
\caption{Typical dependence of the maximal value $h^{\rm max}$ of entropy rate as well as of its mean value $h^{\rm AUC}$ in the range $[0.5 - 2.5]$s for the mHR (blue - m) and the fHR (pink - f) across the cohort, as well as in each subgroup (stressed or control, female or male foetus).
}
\label{fig:complexities_boxplots_no_conditioning}
\end{figure}

We represent how the maximal value $h^{\rm max}$ of the entropy rate and its mean value $h^{\rm AUC}$ for both the mHR and fHR are distributed
 in Fig.~\ref{fig:complexities_boxplots_no_conditioning}. When the entire signal is considered ---~{\em i.e.}, when no conditioning on accelerations or decelerations is performed~---, the entropy rate seems to always have a very slightly higher value in the mHR than in the fHR; this is observed for all subgroups (stressed and control, as well as male or female foetus).

\begin{figure}[htb]
\begin{center}
\includegraphics[width=.49\linewidth]{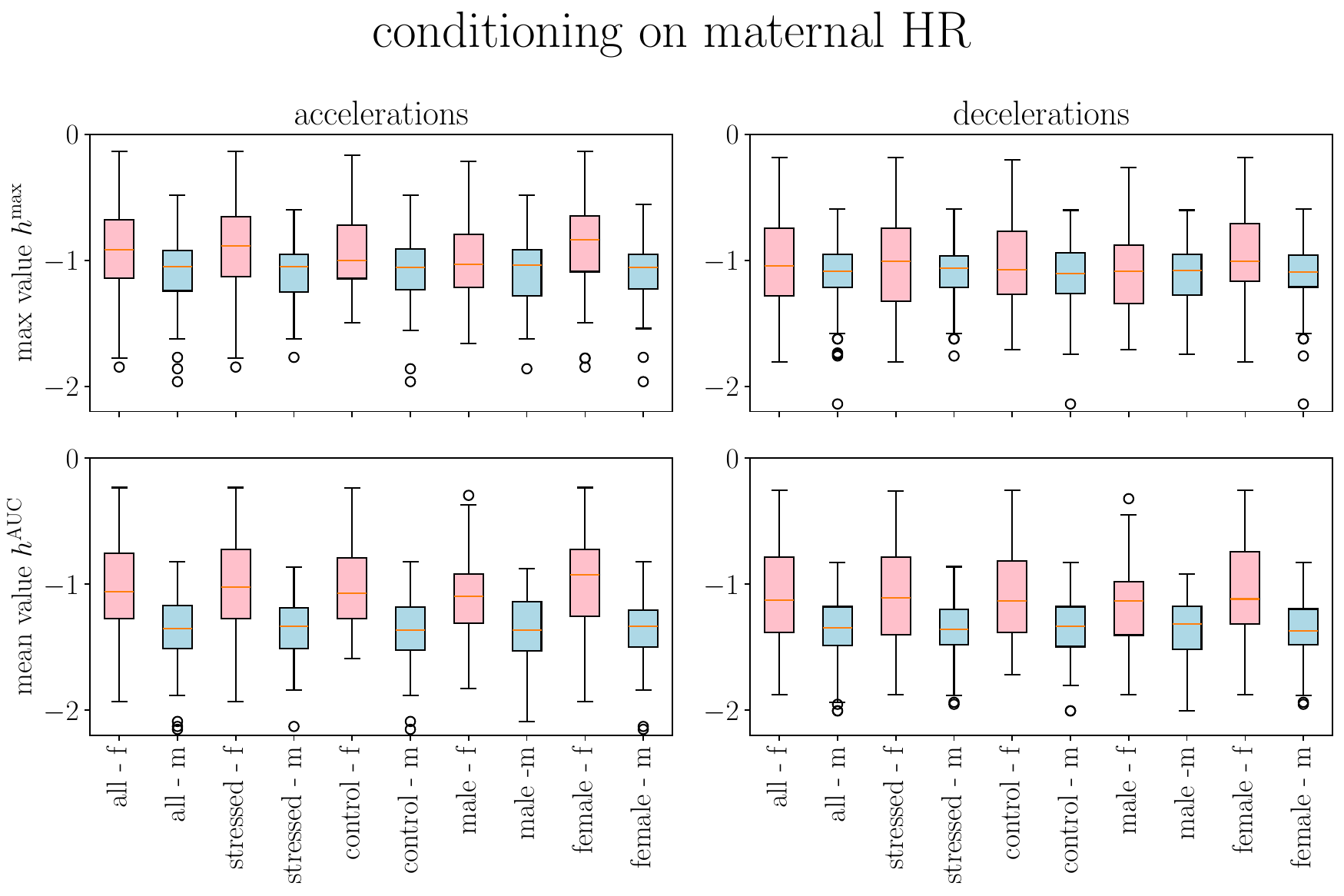}
\includegraphics[width=.49\linewidth]{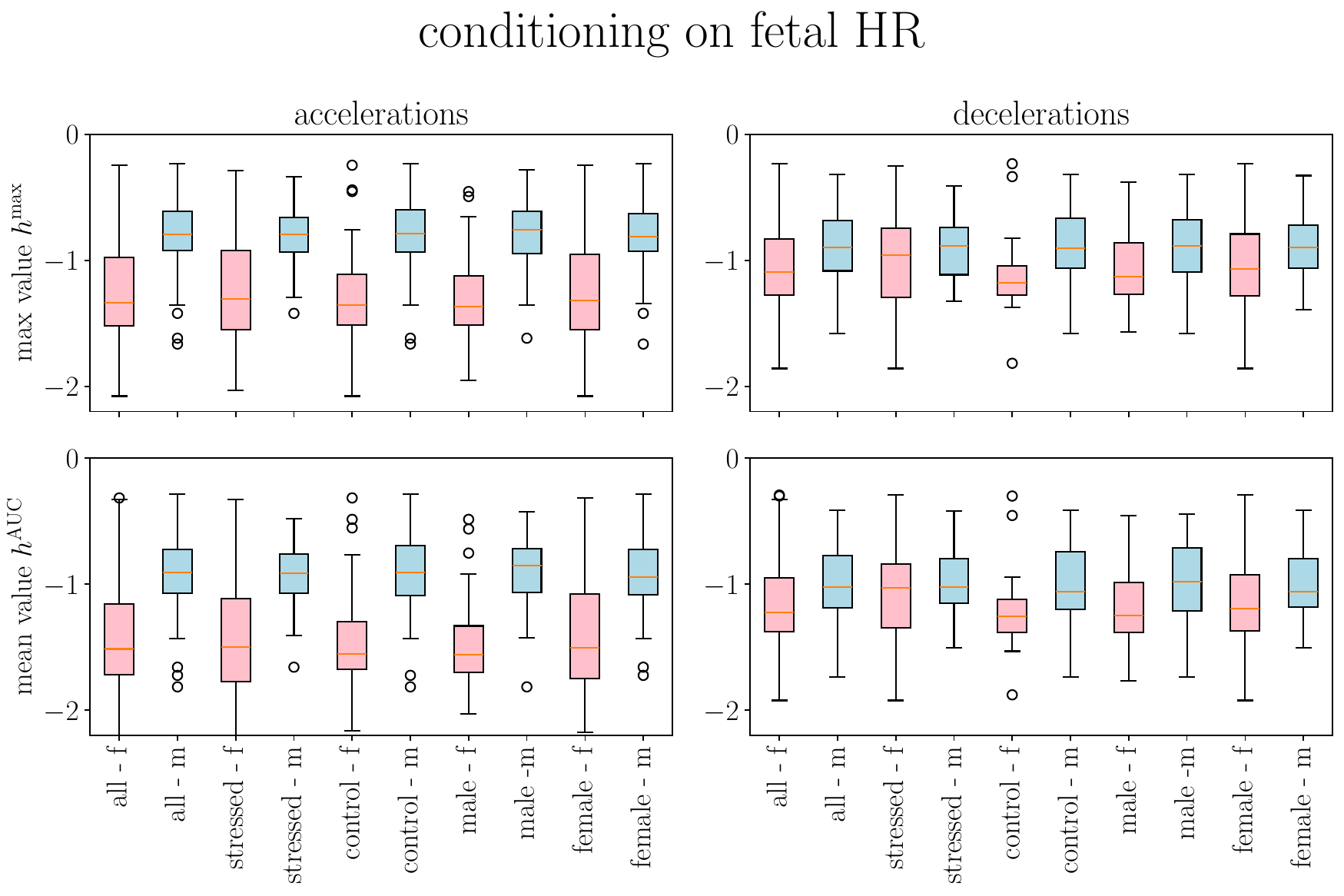}
\end{center}
\caption{Same as Fig.~\ref{fig:complexities_boxplots_no_conditioning} but when considering either accelerations of decelerations, computed on either maternal or fetal HR.
}
\label{fig:complexities_boxplots}
\end{figure}

We perform the same analysis on the conditioned signals by examining mHR and fHR computed in either fetal or maternal accelerations or decelerations.
Boxplots are represented in Fig.~\ref{fig:complexities_boxplots}. Sample entropy box plots are not shown for brevity. The findings are similar to the entropy rate.

Mixed linear model analysis of entropy rate with conditioning framework revealed both univariate signal properties and bivariate maternal-fetal coupling (Figures ~\ref{fig:complexities_boxplots_no_conditioning},~\ref{fig:complexities_boxplots}, Table \ref{tab:mlm_er_complete} ).

\medskip

\textbf{Conditioning Effects: Coupling Strength Quantification}

The magnitude of maternal-fetal coupling can be derived from the MLM $\beta$ coefficients (see \ref{sec:coupling_box} for details). The univariate baseline (no conditioning) represents fetal HR entropy when measured independently ($\beta$ = +0.206, p < 0.001 relative to the reference level of fetal acceleration conditioning). Cross-conditioning on maternal deceleration events reduces entropy by $\beta$ = -0.123 (p = 0.012). The ratio of these coefficients quantifies coupling strength:

\medskip
\textbf{Coupling strength} = $\displaystyle \frac{\mid \beta\,\, {\rm maternal\,\, decelerations}\mid}{\beta \,\, {\rm no\,\, conditioning}} = \frac{0.123}{0.206} = 0.597 \approx 60\%$

\medskip
This indicates that the maternal deceleration coupling effect captures 60\% of the dynamic range established by the no-conditioning baseline. The coupling effect ($\beta$ = -0.123) represents a substantial constraint on fHR complexity—the fHR becomes more predictable and regular during mHR decelerations. This 60\% reduction is STRESS-INVARIANT (p = 0.128 for stress effect on conditioned entropy), indicating it represents a fundamental physiological coordination mechanism present across all mother-fetus pairs, regardless of maternal stress status.

This is the strongest signature of state-dependent bivariate coupling in our results.

We also observed  fetal deceleration conditioning: $\beta$= -0.082, SE = 0.042, p = 0.054, showing a marginal trend toward entropy reduction.

Comparative coupling hierarchy:

\begin{itemize}
    \item Maternal deceleration conditioning: 60\% coupling strength ($\beta$ = -0.123, p = 0.012*)
    \item Fetal deceleration conditioning: 40\% coupling strength ($\beta$ = -0.082, p = 0.054†)
    \item Maternal acceleration conditioning: 16\% coupling strength ($\beta$ = -0.034, p = 0.494, ns)
\end{itemize}

Maternal deceleration events exhibit 1.5$\times$ stronger coupling than fetal events and 3.8$\times$ stronger coupling than maternal accelerations, demonstrating a profound asymmetry in maternal-fetal physiological interdependence. Importantly, this asymmetric pattern is conserved across stressed and control groups, suggesting that it reflects a fundamental maternal-fetal communication architecture rather than a stress-specific adaptation.

The entropy rate and sample entropy exhibited differences in behavior. Hmean was consistently 0.117 units lower than Hmax (Hmean vs Hmax: $\beta$ = -0.117, SE = 0.052, p = 0.023*). This reflects different aspects of entropy rate estimation.

\medskip

\textbf{Group Comparisons: Differential Stress Sensitivity}

We detected no significant demographic effects in conditioned entropy:

\begin{itemize}
    \item Sex (male): $\beta$ = -0.106, p = 0.177
    \item Stress (stressed): $\beta$ = -0.085, p = 0.128 (ns)
    \item Sex $\times$ Stress: $\beta$ = +0.108, p = 0.223 (ns)
\end{itemize}

Of note, while the TE showed significant stress correlations (r = 0.21-0.31 with cortisol, see Table~\ref{tab:te_correlations} ), conditioned entropy rate showed no stress effects (p = 0.128). This differential sensitivity (keeping in mind the exploratory nature of the reported correlation findings) reveals that temporal information transfer (TE) is stress-modulated while the state-dependent coupling (conditioned entropy) is stress-invariant.

\medskip

\subsubsection{Entropy Progression Reveals Coupling Hierarchy}

No conditioning (univariate baseline) \(\leftarrow\) Highest entropy\\
\(\downarrow\) Reference: fetal acceleration conditioning\\
\(\downarrow\) \(\beta = -0.082\) \((p = 0.054,\ \text{trend})\)

Fetal deceleration conditioning\\
\(\downarrow\) \(\beta = -0.123\) \((p = 0.012^*)\)

Maternal deceleration conditioning \(\leftarrow\) Lowest entropy (strongest coupling)

\medskip

The progressive entropy reduction from univariate (no conditioning) to cross-conditioned demonstrates:

\begin{enumerate}
    \item Conditioning constrains signal complexity (reduces entropy);
    \item Cross-conditioning on the OTHER signal's events reveals bivariate coupling;
    \item Maternal deceleration events exert the strongest influence on fetal HR predictability. The lowest entropy indicates here the strongest coupling.
\end{enumerate}

\subsubsection{Sample Entropy MLM Analysis: Data Quality Limitations}

We attempted to assess whether sample entropy exhibited bivariate coupling signatures similar to those of the entropy rate using the same MLM conditioning framework. However, sample entropy values were predominantly zero (87-100\%) in conditioned windows (accelerations/decelerations), reflecting the algorithm's requirement for minimum data points that was not met in brief event windows (Table \ref{tab:er_se_data_quality}).

The data sparsity in conditioned sample entropy values precluded meaningful assessment of bivariate coupling signatures. Sample entropy requires \~100-200 data points for reliable estimation (embedding dimension m=2), but typical acceleration/deceleration events last only 2-10 seconds (8-40 samples at 4 Hz sampling rate), causing the algorithm to fail silently and return zeros.

With only 7-16 non-zero observations per conditioning type (vs hundreds for entropy rate), the simplified MLM analysis (n=286 observations, 120 patients) did not detect significant coupling effects, though effect direction was consistent with entropy rate:

We found evidence of a significant baseline effect (difference from zero), i.e., with no conditioning, sample entropy computed on complete HR data: $\beta$ = +0.971, SE = 0.073, p < 0.001, but no evidence of conditioning on maternal decelerations having on effect on sample entropy reduction: $\beta$ = -0.112, SE = 0.115, p = 0.328 (ns). Meanwhile, the entropy rate showed the same effect ($\beta$ = -0.123), but was significant (p = 0.012), because it had adequate non-zero observations.

The significant baseline (p < 0.001) confirms that sample entropy works on full-time series. The non-significant conditioning effect (p = 0.328) reflects data sparsity, preventing detection of coupling, not absence of coupling. The consistent effect direction (both \~-0.12) with entropy rate suggests the coupling exists but cannot be reliably measured with current data quality.

\subsubsection{Exploratory Neurodevelopmental Associations}

Two SE associations with neurodevelopmental outcomes reached nominal significance (p < 0.05, uncorrected; Table~\ref{tab:erse_correlations}). Neither survived FDR correction (both q = 0.62), consistent with the overall pattern that none of the 144 correlation tests survived multiple comparison adjustment (Fig.~\ref{fig:heatmap_all}).

\textbf{Exploratory interpretation}: Sample entropy during fetal HR accelerations showed tentative positive associations with language receptive scores (higher entropy $\to$ better receptive language). However, failure to survive FDR correction (q = 0.62) indicates these patterns require replication before drawing biological conclusions.

No significant associations were observed between ER/SE features and maternal stress measures (cortisol, PSS, PDQ), tentatively suggesting distinct pathways whereby TE may capture stress physiology while ER/SE relate to neurodevelopment. However, this dissociation also requires replication, given the overall null FDR-corrected results.

\begin{table}
\centering
\caption{Sample entropy associations with neurodevelopmental outcomes (exploratory)}
\label{tab:erse_correlations}
\begin{tabular}{l l l l l l l}\toprule
Feature & Outcome & r & p & q (FDR) & Method & n \\
\midrule
SE fetus (fHR accel) & Language Receptive & +0.290 & 0.026 & 0.62 & Spearman & 59 \\
SE mother (fHR accel) & Language Receptive & +0.290 & 0.026 & 0.62 & Spearman & 59 \\
\bottomrule
\end{tabular}
\begin{flushleft}
\small\textit{Note.} SE = sample entropy; ER = entropy rate; mHR = maternal heart rate; fHR = fetal heart rate; accel = acceleration epochs; decel = deceleration epochs. Both q-values = 0.62, indicating failure to survive FDR correction. No ER features reached p < 0.05 in the overall sample. Previously reported ER/SE-Bayley associations from stratified analyses (by sex or stress group) are not included in this table as they represent post-hoc subgroup findings with even greater multiple comparison burden.
\end{flushleft}
\end{table}

\textbf{Sex-Stratified ER/SE Patterns (Exploratory)}

Sex stratification revealed differential ER/SE association patterns (Fig. \ref{fig:se_er_sex_heatmap}):

\textbf{Female fetuses} (n=71): In addition to the 16 TE correlations (Section 3.3.3), females showed:

\begin{itemize}
    \item ER $\times$ Motor Gross (4 negative correlations, r = -0.37 to -0.61): Fetal ER during various conditioning types showed negative associations with motor gross skills
    \item SE $\times$ Cognitive/Language (3 positive correlations):
    \begin{itemize}
        \item SE mother full $\times$ Cognitive: r = +0.44, p = 0.011, n = 33
        \item SE fetus (fHR accel) $\times$ Lang Receptive: r = +0.41, p = 0.024, n = 30
        \item SE mother (fHR accel) $\times$ Lang Receptive: r = +0.41, p = 0.024, n = 30
    \end{itemize}
\end{itemize}
\textbf{Male fetuses} (n=49): Only the previously noted ER $\times$ Motor Composite correlation (r = +0.39, p = 0.035).

\textbf{Sex $\times$ Stress stratification} revealed additional complexity (Fig. \ref{fig:se_er_sex_stress_heatmap}):

\textbf{Male-Control}: Strong positive ER/SE $\times$ Motor associations dominated (8 ER correlations, 2 SE correlations with r = +0.49 to +0.72), with the strongest being SE fetus full $\times$ Motor Fine (r = +0.72, p < 0.001, $n=20$).

\textbf{Female-Control}: Positive ER $\times$ Cognitive/PDQ associations (5 correlations, r = +0.38 to +0.65) rather than motor associations.

\textbf{Exploratory interpretation}: These sex-stratified patterns (all failing FDR correction with q > 0.40) tentatively suggest sex-differentiated developmental pathways: males show stronger ER/SE-motor coupling (especially in the control subgroup), while females show broader associations spanning cognitive, language, and motor domains. The strong positive SE-motor associations in Male-Control (r = +0.72) contrast with the minimal ER/SE associations in males overall, suggesting stress may disrupt these pathways. All patterns require replication in larger samples.

\clearpage
\subsection{Bivariate information measures of information flow between fHR and mHR: Transfer entropy family}

\subsubsection{TE and conditioning}

\begin{figure}[htb]
\begin{center}
\includegraphics[width=.9\linewidth]{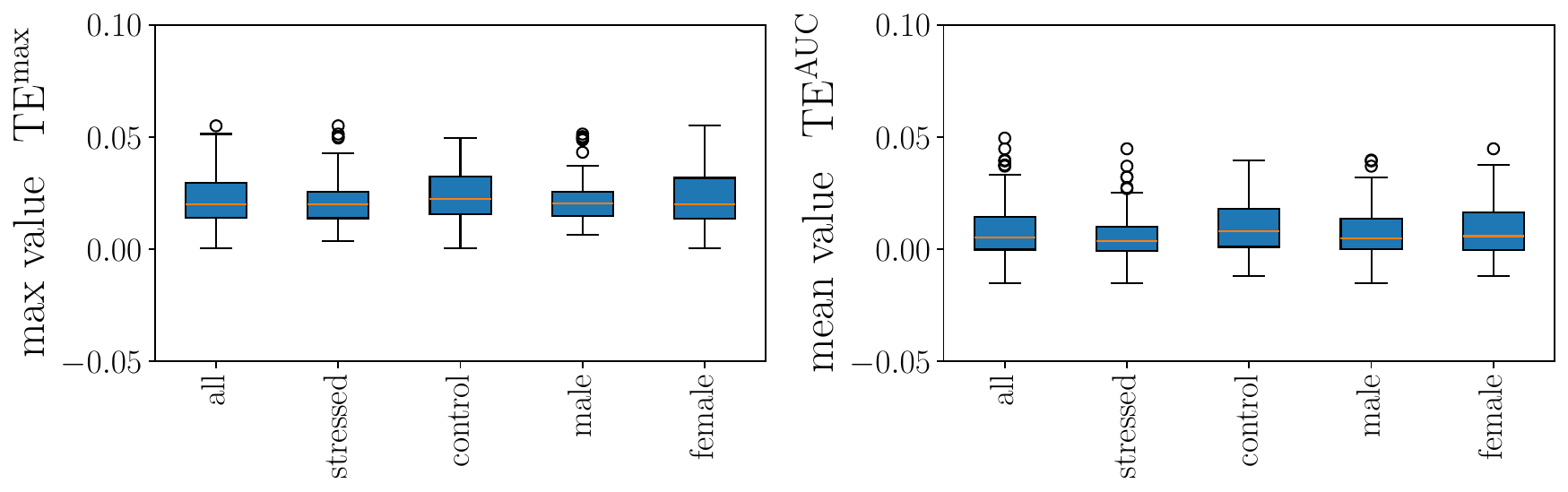}
\end{center}
\caption{
Typical dependence of the maximal value $TE^{\rm max}$ of the net transfer entropy from mother to fetus, as well as of its mean value $TE^{\rm AUC}$ in the range $[0.5 - 2.5]$s across the cohort (all), as well as in each subgroup (stressed or control, female or male fetus). On the left subplots is also presented the components $TE_{m \rightarrow f}$ (mother to fetus) and $TE_{m \rightarrow f}$ (fetus to mother) to show their relative order of magnitude: the net TE is smaller than its one-way components.
}
\label{fig:TE_boxplots_no_conditioning}
\end{figure}

\begin{figure}[htb]
\begin{center}
\includegraphics[width=.9\linewidth]{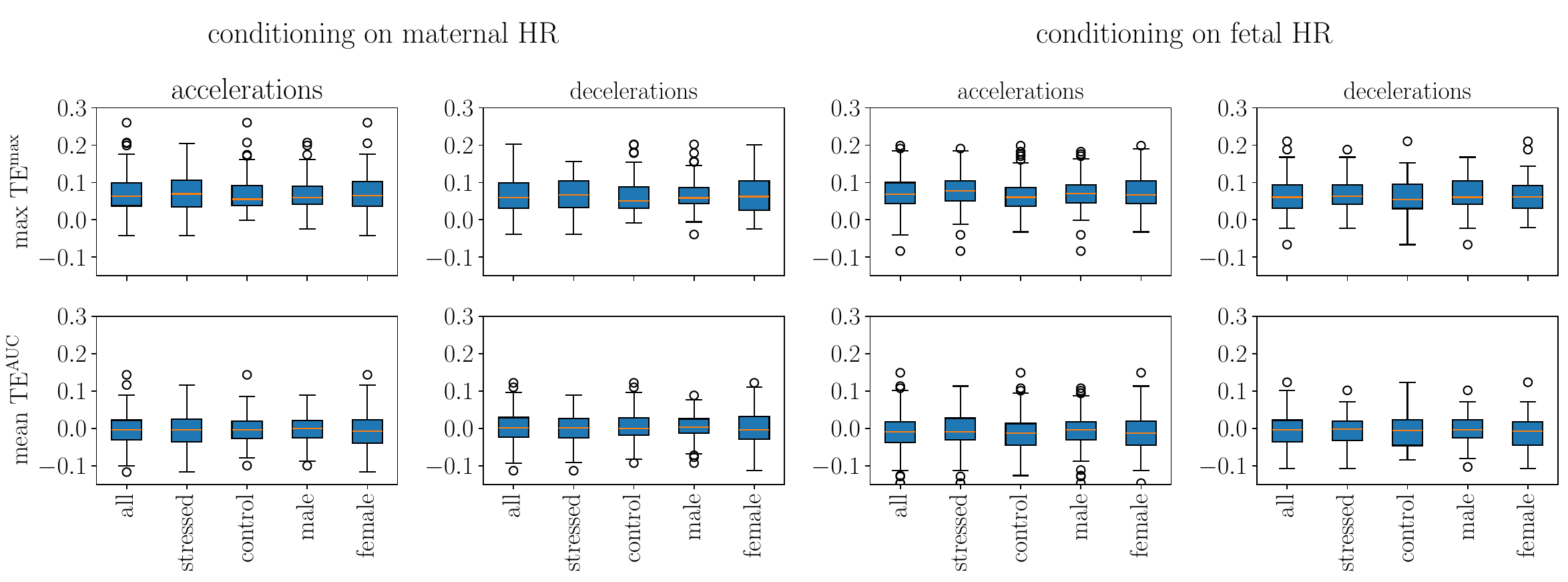}
\end{center}
\caption{Same as Fig.~\ref{fig:TE_boxplots_no_conditioning} for net TE between the mother and the fetus, but when considering either accelerations of decelerations, computed on either maternal or fetal HR.
}
\label{fig:TE_boxplots}
\end{figure}

In this section, we use the time-scale range $[0.5, 2.5]$s and we compute the maximal values, as well as the mean values in this range.
We plot in Fig.~\ref{fig:TE_boxplots_no_conditioning} our findings when using all available points and in Fig.~\ref{fig:TE_boxplots} when using only accelerations, resp. decelerations, each of which is computed using either the maternal HR or the fetal HR. The corresponding statistics for the net TE in Fig.~\ref{fig:TE_boxplots_no_conditioning} and Fig.~\ref{fig:TE_boxplots} are given in Table~\ref{tab:TE_pvalues}; they are obtained from a 1-sided one-sample T-test (positive value versus 0-value) as well as in Table \ref{tab:te_mlm_complete}.

We first comment on the choice of the range of time-scales [0.5, 2.5]s: as can be seen in Fig.~\ref{fig:varying_fs:average}, this corresponds to the range of time-scales where the net TE is expected to be positive, revealing a net information flow from the mother to the fetus.
Looking at Fig.~\ref{fig:TE_boxplots_no_conditioning}, we then comment on the typical values of the net TE, in contrast to the typical values of TE$_{m\rightarrow f}$ between the maternal and the fetal HR and its counterpart TE$_{f\rightarrow m}$ between fetal and maternal HR. The one-way information flow from the mother to the fetus is roughly equivalent to the one-way information flow from the fetus to the mother, and the net TE, which expresses the balance between the two, is much smaller: the net flow of information is small, but directed from the mother to the fetus.

Fig.~\ref{fig:TE_boxplots_no_conditioning} shows that when all available points are used, $TE^{\rm max}$ is slightly positive when considered over the cohort; conversely, $TE^{\rm AUC}$ is indistinguishable from 0. We conclude that
only a small quantity of information
is flowing from the mother to the fetus.

The same observation can be made when any conditioning is applied: $TE^{\rm AUC}$ does not measure any information transfer, while $TE^{\rm max}$ measures a significant information flow from the mother to the fetus (Fig.~\ref{fig:TE_boxplots} and Table~\ref{tab:TE_pvalues}) .
We also observe that max values $TE^{\rm max}$ are larger if a conditioning (on accelerations or decelerations) is used (Fig.~\ref{fig:TE_boxplots}).

\begin{table}[htb]
\centering
\caption{Significance of (a non-vanishing) net transfer entropy ($p$-values)}
\label{tab:TE_pvalues}
\begin{tabular}{@{}rc|ccccc|ccccc@{}}
\toprule
\multirow{3}{*} & & \multicolumn{5}{c}{TE$^{\rm max}$} & \multicolumn{5}{c}{TE$^{\rm AUC}$} \\
      & & & mHR & mHR & fHR & fHR & & mHR & mHR & fHR & fHR \\
 group  & $n$ & all & accel. & decel. & accel. & decel. & all & accel. & decel. & accel. & decel. \\
\midrule
all	    & 120 & 0.18   & 0.089 & 0.085 & 0.074 & 0.082 & 0.25 & 0.53 & 0.50 & 0.57 & 0.56 \\
stressed  &  58 & {\bf 0.031}  & 0.099 & 0.082 & 0.071 & 0.074 & 0.27 & 0.55 & 0.53 & 0.58 & 0.56 \\
control	&  59 & {\bf 0.024}  & 0.082 & 0.085 & 0.074 & 0.092 & 0.23 & 0.51 & 0.47 & 0.56 & 0.55 \\
male	&  49 & {\bf 0.021}  & 0.079 & 0.076 & 0.085 & 0.085 & 0.26 & 0.52 & 0.48 & 0.54 & 0.52 \\
female  &  68 & {\bf 0.032}  & 0.095 & 0.089 & 0.066 & 0.082 & 0.24 & 0.53 & 0.50 & 0.59 & 0.58 \\
\bottomrule
\end{tabular}
\begin{flushleft}
\small\textit{Note.} These are obtained from a 1-sided one sample T-test.
{\em all} refers to values obtained when using all the cohort (stress/control).
{\em stressed}, resp. {\em control}, refers to values obtained in the stressed, resp. control, subgroup.
{\em male}, resp. {\em female}, refers to values obtained in the male fetuses, resp. female fetuses, subgroup.
\end{flushleft}
\end{table}

\medskip

Next, we apply mixed linear model analysis of transfer entropy values across conditioning types. This approach revealed significant effects of conditioning, TE metric type, and demographic interactions (see Table \ref{tab:te_mlm_complete} for details).

\medskip

\textbf{Conditioning and Metric Effects}

We detected significant main effects as follows:

\begin{itemize}
    \item \textbf{TE metric type} (AUC vs Max):  $\beta$ = -0.077, SE = 0.007, p < 0.001*** \\
    TE$^{\rm AUC}$ is consistently 0.077 units lower than TE$^{\rm max}$ across all conditions.
    This reflects, intuitively, the different aspects of information transfer quantification via these two quantities.
    \item \textbf{"Baseline" conditioning} (i.e., none):  $\beta$ = -0.037, SE = 0.007, p < 0.001***\\
    This represents TE when computed on complete time series, without event-specific conditioning
    It serves as a reference for conditioned states (accelerations, decelerations).
\end{itemize}

Importantly, we detected a significant interaction:
\begin{itemize}
    \item \textbf{TE metric $\times$ Conditioning interaction}:  $\beta$ = +0.064, SE = 0.010, p < 0.001***
\end{itemize}
This indicates the difference between mean TE$^{\rm AUC}$ and max TE$^{\rm max}$ varies across conditioning types.
Unconditioned state (none) shows a larger Mean-Max differential than event-conditioned states.

\medskip

\textbf{Sex-Stress Interactions}

Unlike the entropy rate, which showed no demographic modulation, TE revealed significant sex-stress interactions:

\begin{itemize}
\item \textbf{Stress (main effect)}: $\beta$ = +0.023, SE = 0.010, p = 0.026*\\
    The stressed group shows higher TE values overall.
    This is consistent with the here-reported, albeit exploratory, TE-cortisol correlations.

\item \textbf{Sex $\times$ Stress interaction}: $\beta$ = -0.042, SE = 0.016, p = 0.009**\\
    The effect of stress differs between male and female fetuses.
    Male stressed fetuses show lower TE than expected from additive effects.

\item \textbf{Sex $\times$ Stress $\times$ Conditioning interaction}: $\beta$ = +0.037, SE = 0.015, p = 0.014*\\
    Three-way interaction indicates that sex-stress differences vary across conditioning types.
    Most pronounced in the unconditioned (baseline) state.
\end{itemize}

\subsubsection{Differential Stress Sensitivity}

Transfer entropy shows significant stress modulation (p = 0.026) and sex-stress interactions (p = 0.009), contrasting sharply with entropy rate which showed no stress effects (p = 0.128).

This differential sensitivity reveals distinct physiological mechanisms:

\begin{itemize}
    \item Transfer Entropy: Stress-sensitive (p = 0.026) - Temporal information flow modulated by maternal stress
    \item Conditioned Entropy Rate: Stress-invariant (p = 0.128) - State-dependent coupling robust to stress
\end{itemize}

\textbf{Biological interpretation}: Stress influences temporal prediction dynamics (how maternal past predicts fetal future) but not instantaneous state dependencies (how concurrent maternal states constrain fetal complexity). This suggests:

\begin{itemize}
    \item Stress-modulated neural or hormonal pathways affect lagged predictive relationships (TE)
    \item Fundamental autonomic coordination mechanisms remain robust to acute stress (conditioned entropy)
\end{itemize}

The significant sex-stress interaction in TE ($\beta$ = -0.042, p = 0.009) indicates that stress effects on temporal coupling differ between male and female fetuses, potentially reflecting sex-specific stress-response pathways or differences in autonomic regulation in utero.

\subsubsection{Exploratory Neurodevelopmental Associations of TE}

Of 144 correlation tests performed across all entropy features and outcomes, seven uncorrected associations ($p < 0.05$) were identified, matching the expected false positive rate of $\sim 7.2$ at $\alpha$ = 0.05 (Table~\ref{tab:te_correlations} and Fig.~\ref{fig:heatmap_all}). Critically, none of these associations survived false discovery rate (FDR) correction (all q > 0.40), indicating these findings are exploratory and require independent replication before biological interpretation.

\textbf{TE-Cortisol associations}: Five TE-cortisol associations reached nominal significance (p < 0.05, uncorrected; Table~\ref{tab:te_correlations} ).

\begin{table}
\centering
\caption{Transfer entropy associations with maternal cortisol (exploratory)}
\label{tab:te_correlations}
\begin{tabular}{l l l l l }\toprule
Feature & r & p & q (FDR) & n \\\midrule

Max TE (mHR conditioning, decel) & +0.315 & 0.003 & 0.41 & 88 \\
Max TE (mHR conditioning, accel) & +0.287 & 0.007 & 0.48 & 88 \\
Max TE (fHR conditioning, decel) & +0.271 & 0.011 & 0.51 & 88 \\
Max TE (fHR conditioning, accel) & +0.250 & 0.019 & 0.62 & 88 \\
Mean TE (mHR conditioning, accel) & +0.221 & 0.038 & 0.68 & 88 \\ \bottomrule

\end{tabular}
\begin{flushleft}
\small\textit{Note.} All correlations were computed using the Spearman method. TE = transfer entropy; mHR = maternal heart rate; fHR = fetal heart rate. All q-values > 0.40 indicate failure to survive FDR correction. No other TE-cortisol associations reached $p < 0.05$.
\end{flushleft}
\end{table}

\textbf{Exploratory interpretation}: All uncorrected significant TE-cortisol associations were positive (r = 0.22-0.31), tentatively suggesting that stronger directional coupling between maternal and fetal heart rates may be associated with higher chronic stress. Maximum TE values showed stronger associations than mean TE values. However, given the failure to survive multiple comparison correction, these patterns require replication.

\textbf{TE-Bayley association}: One TE-neurodevelopmental association reached nominal significance (Table \ref{tab:te_bayley_correlations}).

\begin{table}
\centering
\caption{Transfer entropy association with neurodevelopment (exploratory)}
\label{tab:te_bayley_correlations}

\begin{tabular}{l l l l l l }\toprule
Feature & Outcome & r & p & q (FDR) & n \\\midrule

Max TE (fHR conditioning, accel) & Language Expressive & -0.277 & 0.036 & 0.73 & 58 \\ \bottomrule

\end{tabular}
\begin{flushleft}
\small\textit{Note.} All correlations were computed using the Spearman method. TE = transfer entropy; fHR = fetal heart rate. q = 0.73 indicates failure to survive FDR correction. No other TE-Bayley associations reached $p < 0.05$.
\end{flushleft}
\end{table}

\textbf{Exploratory interpretation}: Higher TE during fetal accelerations was tentatively associated with lower expressive language skills. This was the only TE-neurodevelopmental association to reach nominal significance among 30 tests performed, and it did not survive FDR correction (q = 0.73), suggesting possible Type I error. No significant TE associations were observed with cognitive or motor outcomes, or with psychological stress measures (PSS, PDQ).

\subsubsection{Sex-stratified TE analysis (Exploratory)}

Sex-stratified analysis revealed striking differences in TE-outcome associations, consistent with the robust Sex $\times$ Stress $\times$ TE interaction from MLM analysis:

\textbf{Female fetuses }(n=71) exhibited 16 significant correlations (p < 0.05, uncorrected), but none survived FDR (all q > 0.40).

Specifically, we observed four TE-Cortisol correlations:

\begin{itemize}
    \item Max TE (fHR accel): r = +0.368, p = 0.007, n = 53
    \item Max TE (fHR decel): r = +0.373, p = 0.006, n = 53
    \item Max TE (mHR accel): r = +0.351, p = 0.009, n = 54
    \item Max TE (mHR decel): r = +0.423, p = 0.002, n = 53
\end{itemize}

We also observed five negative TE-Cognitive performance correlations:

\begin{itemize}
    \item Max TE (fHR all): r = -0.410, p = 0.018, n = 33
    \item Max TE (fHR accel): r = -0.401, p = 0.023, n = 32
    \item Max TE (mHR all): r = -0.410, p = 0.018, n = 33
    \item Mean TE (fHR all): r = -0.377, p = 0.031, n = 33
    \item Mean TE (mHR all): r = -0.377, p = 0.031, n = 33
\end{itemize}

\textbf{Male fetuses} (n=49) exhibited only one significant correlation (Entropy Rate fetus mHR decel $\times$ Motor Composite: r = +0.387, p = 0.035, n = 30), which did not survive FDR correction (q > 0.40). No significant TE-cortisol or TE-cognitive performance correlations were seen in males.

\textbf{Exploratory interpretation}: Female-specific maternal-fetal coupling mechanisms may underlie the robust Sex $\times$ Stress $\times$ TE interaction. In females, higher TE tentatively associates with both higher maternal stress (cortisol) and lower cognitive scores, suggesting that increased directional coupling under stress may reflect compensatory or maladaptive physiological responses. The absence of TE correlations in males tentatively suggests sex-differentiated autonomic regulation pathways. However, all correlation findings failed FDR correction and require replication in adequately powered studies.

\subsubsection{Sex $\times$ Stress Interaction Patterns (Exploratory)}

Further stratification by sex and stress status revealed complex interaction patterns (Figure \ref{fig:se_er_sex_stress_heatmap}):

Male-Control fetuses (n=30) exhibited 16 significant correlations (p < 0.05, uncorrected; all q > 0.40):

\begin{itemize}
    \item Strong positive ER/SE $\times$ Motor associations (r = +0.49 to +0.72)
    \item Positive Mean TE $\times$ PSS associations (r = +0.43 to +0.45)
    \item Positive Mean TE $\times$ Cognitive (r = +0.48)
\end{itemize}

Male-Stressed fetuses (n=19) showed seven significant correlations (all q > 0.40):

\begin{itemize}
    \item Pattern reversal: Negative Max TE $\times$ PSS (r = -0.49, p = 0.032)
    \item Negative Mean TE $\times$ Language scores (r = -0.65 to -0.70)
    \item Motor associations absent
\end{itemize}

Female-Control fetuses (n=32) exhibited twelve significant correlations (all q > 0.40):

\begin{itemize}
    \item Strongest TE-cortisol coupling: Max TE (mHR decel) r = +0.55, p = 0.006
    \item Very strong negative TE $\times$ Language: Max TE (fHR accel) $\times$ Lang Composite r = -0.86, p < 0.001
    \item Positive ER $\times$ Cognitive/PDQ (r = +0.38 to +0.65)
\end{itemize}

Female-Stressed fetuses (n=39) showed seven significant correlations (all q > 0.40):

\begin{itemize}
    \item Weaker TE-cortisol coupling (r = +0.38 vs r = +0.55 in control)
    \item Shift from TE to SE/ER dominance for cognitive/language outcomes
    \item Positive SE mother $\times$ Cognitive (r = +0.54, p = 0.014)
\end{itemize}

\textbf{Exploratory interpretation}: These sex $\times$ stress patterns (all requiring replication due to FDR failure) tentatively suggest that in:

\begin{enumerate}
    \item \textbf{Males}: Stress eliminates positive motor associations and reverses the coupling-stress relationship direction;
    \item \textbf{Females}: Stress weakens maternal-fetal coupling but activates alternative signal complexity pathways (SE/ER).

\end{enumerate}

These exploratory stratified findings provide potential mechanistic context for the robust Sex $\times$ Stress $\times$ TE interaction, though the specific correlation patterns require validation in larger samples.

\subsection{Multivariate Modeling: Sample Size Limitations Prevent Reliable Inference}

Multivariate analysis examined whether entropy features jointly predict neurodevelopmental outcomes using regularized regression and dimensionality reduction approaches designed for high-dimensional data. However, severe sample-size limitations relative to the number of predictors (52 entropy features + a stress indicator) prevented reliable multivariate inference. The n/k ratio should ideally exceed 10-20 for reliable multivariate inference. The observed ratios of \~1.2 (sample size/predictors ratio) indicate severe underpowering, approximately 10-fold below recommended thresholds.

\subsubsection{Multicollinearity Assessment}

Variance Inflation Factor analysis revealed severe multicollinearity among entropy features. Across outcomes, 91--94\% of features (48--50 of 53) had VIF $>$ > 10.

Highest VIF features (VIF > 100):

\begin{itemize}
    \item \verb|max_TE_mHR_all|, \verb|max_TE_fHR_all|, \verb|mean_TE_mHR_all| (TE features computed on "all" time points)
    \item These TE "all" features are mathematically related, explaining the extreme multicollinearity
\end{itemize}

This indicates severe redundancy that necessitates regularization approaches but also reflects fundamental mathematical relationships between features computed from the same time series.

\subsubsection{Model Performance}

Cross-validated $R^2$ values for multivariate models are presented in Table~\ref{tab:model_comparison}.

\begin{table}
\centering
\caption{Cross-validated $R^2$ by model and outcome}
\label{tab:model_comparison}
\begin{tabular}{l l l l l}
\toprule
Outcome & Elastic Net & PCA+Ridge & PLS & Random Forest \\
\midrule
Cognitive Composite & -0.10 & -0.06 & -0.79 & -0.10 \\
Language Composite & -0.42 & -0.57 & -1.23 & -0.40 \\
Motor Composite & -0.49 & -0.52 & -0.99 & -0.43 \\
Motor Fine Skills & -0.23 & -0.01 & +0.10 & -0.19 \\
Motor Gross Skills & -0.18 & -0.04 & -0.52 & -0.15 \\
\bottomrule
\end{tabular}

\end{table}

Negative CV-$R^2$ values indicate that models performed worse than a null model predicting the sample mean, reflecting overfitting due to the unfavorable ratio of predictors to observations. Only PLS regression for Motor Fine Skills achieved a marginally positive CV-$R^2$ of 0.10, explaining 10\% of the variance. This pattern reflects severe overfitting due to the high dimensionality (53 predictors) relative to the small sample size (n = 30-66).

\subsubsection{Feature Selection Patterns}

\textbf{Elastic Net} selected features are shown by way of an example for the cognitive composite (Table \ref{tab:elastic_net}):

\begin{table}
\centering
\caption{Elastic Net selected features for cognitive performance composite}
\label{tab:elastic_net}
\begin{tabular}{l l}\toprule
Feature & Coefficient \\\midrule

SE\_mother\_mHR\_accel & +0.095 \\
stress\_binary & -0.094 \\
max\_TE\_fHR\_all & -0.067 \\
max\_TE\_mHR\_all & -0.067 \\
SE\_fetus\_mHR\_accel & +0.053 \\ \bottomrule

\end{tabular}

\end{table}

TE features were selected by Elastic Net with moderate coefficients, though model performance remained poor (CV-$R^2$ = -0.10).

\textbf{Random Forest} Feature Importance (Top TE Features, Table \ref{tab:rf_bayley}):

\begin{table}
\centering
\caption{Random Forest top TE features for Bayley test}
\label{tab:rf_bayley}
\begin{tabular}{l l l}\toprule
Outcome & Top TE Features & Importance Score \\\midrule

Cognitive & mean\_TE\_fHR\_all & 0.065 \\
Cognitive & max\_TE\_fHR\_all & 0.061 \\
Motor Composite & max\_TE\_fHR\_all & 0.079 \\
Motor Composite & max\_TE\_mHR\_decel & 0.071 \\
Motor Fine Skills & max\_TE\_mHR\_decel & 0.092 \\ \bottomrule

\end{tabular}

\end{table}

TE features consistently ranked among the top Random Forest predictors, particularly for motor outcomes, despite overall poor model performance.

\subsubsection{PLS Loading Analysis}

Partial least squares regression identified latent components with the highest feature loadings (Table \ref{tab:pls_bayley}):

Top Loading Features by Outcome:

\begin{table}
\centering
\caption{PLS loadings for Bayley test}
\label{tab:pls_bayley}
\begin{tabular}{l l l l}\toprule
Outcome & Feature Type & Feature & Loading \\\midrule

Cognitive & TE & max\_TE\_mHR\_all & 0.430 \\
Cognitive & TE & max\_TE\_fHR\_all & 0.430 \\
Cognitive & TE & mean\_TE\_mHR\_all & 0.416 \\
Language & ER & ER\_mother\_mHR\_accel & 0.364 \\
Language & ER & ER\_mother\_full & 0.349 \\
Motor Composite & ER & ER\_mother\_mHR\_decel & 0.353 \\
Motor Fine Skills & TE & mean\_TE\_mHR\_all & 0.381 \\
Motor Gross Skills & TE & max\_TE\_fHR\_all & 0.367 \\ \bottomrule

\end{tabular}

\end{table}

Pattern: TE features dominated PLS loadings for cognitive and motor outcomes, while ER features dominated for language outcomes, suggesting domain-specific relevance despite poor overall predictive performance.

\subsubsection{Parsimonious Forward Selection Results}

Using a maximum of 3 features to optimize the n/k ratio we found the following statistics:

\textbf{Cognitive Composite}:

\begin{itemize}
    \item Selected: mean\_TE\_mHR\_all, SE\_mother\_mHR\_accel
    \item $R^2$(adj) = 0.067
    \item No significant predictors
\end{itemize}

\textbf{Language Composite}:

\begin{itemize}
    \item Selected: SE\_mother\_mHR\_decel
    \item $R^2$(adj) = 0.116
    \item Stress main effect: $\beta$ = -13.4, p = 0.051 (marginal)
\end{itemize}

\textbf{Motor Composite}:

\begin{itemize}
    \item Selected: max\_TE\_fHR\_all, max\_TE\_mHR\_all, SE\_mother\_fHR\_accel
    \item $R^2$(adj) = 0.082
    \item TE features selected but not individually significant
\end{itemize}

\textbf{Motor Fine Skills}:

\begin{itemize}
    \item Selected: SE\_fetus\_mHR\_accel, max\_TE\_fHR\_all
    \item $R^2$(adj) = 0.007
    \item No significant predictors
\end{itemize}

Even with parsimonious models (maximum three features), predictive performance remained poor, and individual features were not statistically significant, reinforcing the severe underpowering of multivariate approaches in this sample.

\subsubsection{Integration with the univariate findings}

\textbf{TE-Cortisol Pattern Persists}:

While multivariate models failed to achieve reliable predictive performance, the pattern observed in univariate correlations was consistent: TE features were selected by regularized models and ranked highly in feature importance analyses, particularly for motor and cognitive outcomes. However, the lack of direct TE-Bayley correlations (only one significant of 30 tests in univariate analysis) was confirmed by poor multivariate performance.

\textbf{ER/SE Show Domain Specificity}:

PLS loading analysis suggested potential domain specificity: TE features for cognitive/motor outcomes, ER features for language outcomes. However, given the severe underpowering ($n/k \simeq 1.2$) and negative CV-R$^2$ values, these patterns should be considered exploratory hypotheses requiring replication in adequately powered samples.

\subsubsection{Interpretation}

The multivariate analysis with 52 entropy features reveals:

\begin{enumerate}
    \item Severe underpowering ($n/k$ $\simeq$ 1.2 vs optimal >10-20) prevents reliable multivariate inference;
    \item Extreme multicollinearity (91-94\% features with VIF > 10) reflects mathematical relationships between features;
    \item Poor predictive performance (mostly negative CV-$R^2$) indicates overfitting despite regularization;
    \item Feature selection patterns suggest potential relevance of TE for motor/cognitive and ER for language, but lack statistical reliability;
    \item Univariate analyses remain most interpretable given sample size constraints.

\end{enumerate}

In summary, the univariate correlation analyses and the MLM analyses with reduced feature sets provide the most reliable findings for this sample size. Multivariate models would require sample sizes of n > 500 (approximately 10 $\times$ 52 features) for valid inference.

\subsection{Methodological insights: Dependence of TE on HR sampling rate}
\label{sec:sampling_effect}

Acquiring data at the correct sampling rate is crucial for accurate estimation of the HRV metrics. Here, we test the impact of sampling rates on TE estimates, both with and without conditioning on heart rate dynamics (accelerations and decelerations). We show that TE estimated on full HR time series does not depend on sampling rate, but TE of the conditioned HR does.

After the filtering stage, we resampled the data at the frequency $f_s=20$Hz.
To validate this choice of the sampling rate for our information-theoretical approach, we also explored several other values of the sampling rate $f_s$: lower values 4Hz, 10Hz, as well as larger values 100Hz and even 1000Hz corresponding to the ECG sampling rate.

\begin{figure}[htb]
\begin{center}
\includegraphics[width=.9\linewidth]{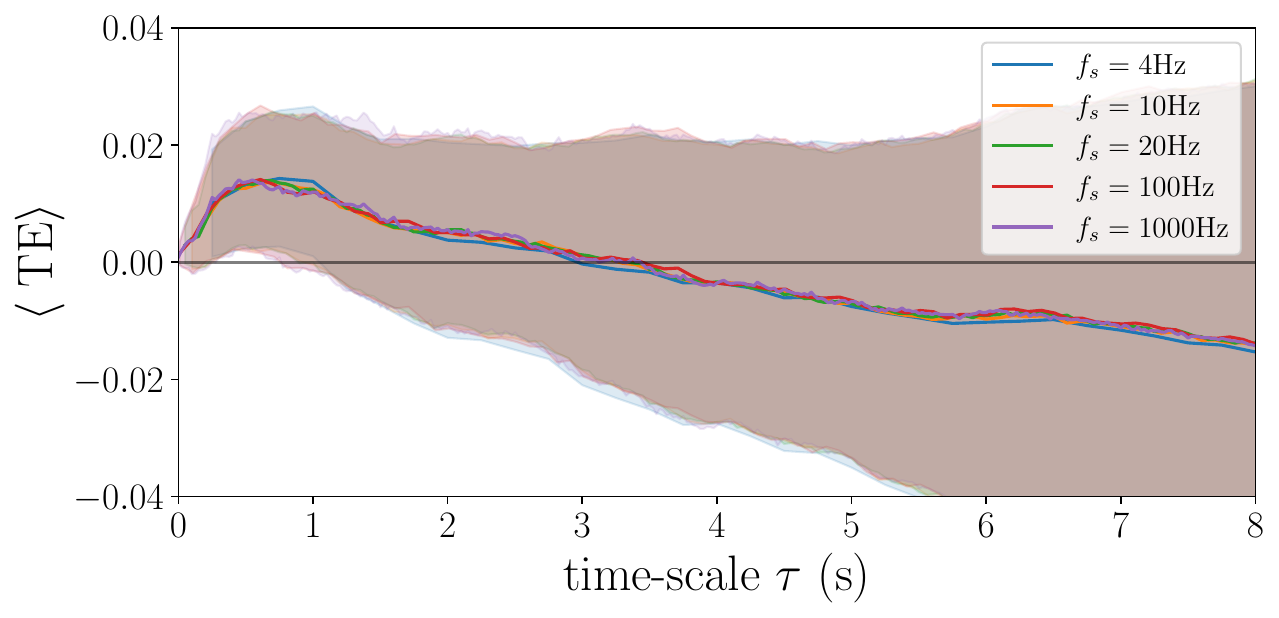}
\end{center}
\caption{Ensemble-averaged $\langle {\rm TE}\rangle$ as a function of the time-scale $\tau$. The shaded areas correspond to the dispersion around the average due to inter-individual variability and are measured as the standard deviation across the cohort.
Both the ensemble-averaged TE and its dispersion do not depend on the sampling frequency.}
\label{fig:varying_fs:average}
\end{figure}

Figure~\ref{fig:varying_fs:average} shows the evolution with the timescale $\tau$ of the ensemble-averaged $\langle {\rm TE}\rangle$, measured by averaging the net TE over the entire cohort. We observe that whatever the sampling frequency $f_s$ used in the pre-processing described in section~\ref{sec:filtering}, the evolution of $\langle {\rm TE}\rangle$ with the timescale $\tau$ is the same.

For all other results in this article, we selected $ f_s=20$ Hz, but we note that 10Hz or even 4Hz is sufficient.


\section{Discussion}

\subsection{Summary of Key Findings}

This study extends our previous work establishing the Fetal Stress Index ~\cite{Lobmaier:2020} by exploring the information-theoretical foundations of maternal-fetal heart rate coupling. We demonstrate that prenatal maternal stress influences specific aspects of physiological communication between mother and fetus, while other fundamental coupling mechanisms remain conserved.

\textbf{Principal findings:}

First, we identified dual coupling mechanisms operating simultaneously during the third trimester: temporal information transfer (quantified by transfer entropy) and state-dependent synchronization (quantified by conditioned entropy). These complementary measures reveal that maternal-fetal coupling involves both time-lagged predictive relationships and concurrent state dependencies.

Second, we discovered differential stress sensitivity in these coupling pathways. While temporal information transfer shows stress-related modulation and exploratory associations with maternal cortisol, state-dependent coupling remains stress-invariant—a fundamental physiological coordination that persists regardless of maternal stress status.

Third, our analyses revealed profound asymmetry in maternal-fetal coupling. Maternal heart rate decelerations exert substantially stronger influence on fetal heart rate complexity than any other physiological state, reducing fetal entropy by approximately 60\%. This asymmetric coupling may reflect critical regulatory states where maternal-fetal coordination is most pronounced.

Finally, exploratory analyses suggested sex-differentiated coupling patterns, with female fetuses exhibiting stronger temporal coupling associations than male fetuses. These sex-specific patterns require replication but align with emerging evidence of sexually dimorphic fetal autonomic development.

We contextualize these findings within the broader literature on prenatal stress programming, maternal-fetal physiological communication, and autonomic nervous system development.

\subsection{From BPRSA to Information Theory: Deepening the Fetal Stress Index Framework}

Our previous work ~\cite{Lobmaier:2020} established the Fetal Stress Index (FSI) using bivariate phase-rectified signal averaging (BPRSA) to quantify maternal-fetal heart rate coupling. That study revealed a fundamental observation: while control fetuses remained physiologically "stable" during maternal breathing cycles, stressed fetuses exhibited fetal heart rate decreases in response to maternal heart rate decreases. The FSI successfully discriminated between stressed and control groups and correlated with maternal perceived stress, validating the concept that maternal stress alters the fetal autonomic response to maternal physiological fluctuations.

The current study builds upon this foundation by applying information-theoretical measures to elucidate the mechanisms underlying these coupling phenomena. Where BPRSA captures phase-rectified signal relationships, transfer entropy quantifies directional information flow—specifically, how the maternal heart rate past improves prediction of the fetal heart rate future beyond the fetal signal's own history. Where BPRSA identifies average signal responses, entropy rate conditioning reveals how signal complexity changes during specific physiological states, offering insights into state-dependent coupling mechanisms.

The ~\cite{Lobmaier:2020} observation that stressed fetuses show heart rate decreases during maternal deceleration events finds theoretical grounding in our conditioning framework. The substantial entropy reduction during maternal decelerations—approximately 60\% compared to baseline—demonstrates that fetal heart rate becomes highly constrained and predictable during these maternal states. This state-dependent synchronization represents the information-theoretical substrate of the coupling captured by BPRSA.

Critically, we extend the FSI framework by demonstrating that coupling operates through dual pathways. The stress-invariant state-dependent coupling (conditioned entropy) may represent the fundamental physiological coordination mechanism that exists across all mother-fetus pairs, while the stress-sensitive temporal coupling (transfer entropy) may capture the "over-sensitization" hypothesized in our previous work—where maternal stress alters the fetal autonomic system's temporal response dynamics to maternal physiological changes.

\subsection{Biological Foundations of Maternal-Fetal Coupling: Literature Context}

\subsubsection{Prenatal Stress Programming and the HPA Axis}

The fetal programming hypothesis posits that prenatal environmental conditions, including maternal psychological stress, can produce lasting alterations in fetal development through multiple pathways ~\cite{Barker1990, Gluckman2008}. Maternal stress activates the hypothalamic-pituitary-adrenal (HPA) axis, elevating cortisol levels that can cross the placental barrier despite partial 11$\beta$-HSD2 enzymatic protection ~\cite{Seckl2004, ODonnell2012}. Our previous finding of 63\% higher hair cortisol concentrations in stressed mothers ~\cite{Lobmaier:2020} confirmed chronic HPA axis activation in our cohort.

Elevated fetal glucocorticoid exposure can program the developing HPA axis and autonomic nervous system, altering set points for physiological regulation ~\cite{Matthews2012, Monk2012}. The exploratory associations between transfer entropy and maternal cortisol observed in our study—while not surviving correction for multiple comparisons and requiring replication—align with this programming framework. They suggest that chronic maternal stress may alter the temporal dynamics of maternal-fetal heart rate coupling, potentially reflecting modified autonomic responsiveness in the developing fetus.

\subsubsection{Autonomic Nervous System Maturation}

The fetal autonomic nervous system undergoes rapid maturation during the third trimester, with progressive increases in parasympathetic tone and heart rate variability ~\cite{Schneider2008, VanLeeuwen2014}. This developmental trajectory is sensitive to environmental perturbations, including maternal stress ~\cite{DiPietro2006}.

Our finding of universal acceleration predominance—stronger in fetal than maternal heart rate and independent of sex or stress—may reflect fundamental developmental constraints on autonomic regulation. The predominance of heart rate accelerations over decelerations in healthy fetuses has been documented previously ~\cite{Dawes1981} and likely represents the balance between sympathetic and parasympathetic influences during normal third-trimester development.

The asymmetric coupling we observed, where maternal decelerations exert a stronger influence on fetal dynamics than maternal accelerations, may relate to the physiological salience of bradycardic events. Maternal heart rate decelerations could signal states requiring heightened maternal-fetal coordination—perhaps related to maternal respiratory patterns, as suggested by our previous work linking maternal breathing cycles to fetal heart rate responses ~\cite{Lobmaier:2020}. The mechanical effects of diaphragm excursion on uterine pressure, combined with associated changes in maternal oxygenation and autonomic balance, may create particularly potent coupling conditions.

\subsubsection{Maternal-Fetal Physiological Communication Pathways}

Multiple pathways can mediate maternal-fetal coupling. Direct mechanical transmission through uterine wall movement and amniotic fluid dynamics can influence fetal heart rate ~\cite{Ohta1999}. Shared placental circulation creates metabolic coupling, with maternal blood gas changes rapidly affecting fetal oxygenation ~\cite{Longo1987}. Maternal autonomic fluctuations can alter uterine blood flow, indirectly influencing fetal cardiovascular regulation ~\cite{Metsala1993}.

Our distinction between stress-sensitive temporal coupling and stress-invariant state-dependent coupling may reflect different communication pathways. The state-dependent coupling—robust across stress conditions and demographic variation—could represent direct mechanical or circulatory coupling that remains constant. The temporal coupling—modulated by stress and showing sex-specific patterns—might involve more complex autonomic and hormonal pathways susceptible to maternal stress effects.

This interpretation aligns with the "fetal stress memory" concept proposed in our previous work ~\cite{Lobmaier:2020}, in which stressed fetuses exhibited altered responses, suggesting persistent programming effects. The information-theoretical framework reveals that this programming may specifically affect temporal prediction dynamics while preserving fundamental state-dependent coordination mechanisms.

\subsection{Sex Differences in Fetal Stress Responses: Exploratory Observations}

Sex differences in prenatal development and stress vulnerability are increasingly recognized ~\cite{Clifton2010, Bale2016}. Male fetuses typically grow faster but may be more vulnerable to adverse prenatal conditions, while female fetuses show more adaptive physiological responses to environmental challenges ~\cite{Eriksson2010, Sandman2013}.

Our exploratory sex-stratified analyses revealed striking patterns, though none survived correction for multiple comparisons and all require independent replication. Female fetuses showed numerous transfer entropy associations with maternal cortisol and neurodevelopmental outcomes that were entirely absent in males. Male fetuses instead showed entropy rate and sample entropy associations with motor development, particularly in the control subgroup, which disappeared under maternal stress conditions.

These tentative patterns align with the hypothesis that female fetuses may exhibit more pronounced maternal-fetal physiological coupling, potentially as an adaptive mechanism for monitoring maternal stress ~\cite{Buss2009}. Male fetuses may rely more on signal complexity measures related to their own autonomic maturation, with maternal stress disrupting these developmental trajectories.

The robust sex-by-stress interaction in transfer entropy from our mixed linear model analysis provides the only statistically reliable evidence for sex-differentiated coupling mechanisms in our dataset. This finding warrants mechanistic investigation in adequately powered studies designed specifically to test sex-specific hypotheses.

Several biological mechanisms could underlie sexually dimorphic stress responses. Differential placental function by fetal sex has been documented, with female placentas showing more adaptive responses to maternal stress through altered gene expression and metabolic profiles ~\cite{Sandman2013, Rosenfeld2015}. Sex steroid hormones, present even in fetal life, can modulate HPA axis and autonomic nervous system development differently in males and females ~\cite{Bale2016}. The timing of autonomic nervous system maturation differs between sexes, potentially creating windows of differential vulnerability ~\cite{DiPietro2004}.

\subsection{Maternal-Fetal Coupling: Mechanisms, Quantification and Clinical Implications}

\subsubsection{Why "Coupling"? Defining Physiological Interdependence}

Before interpreting the asymmetric patterns we observed, it is essential to clarify what we mean by "coupling" and why this term appropriately describes our findings.

In physiological systems, coupling refers to the condition where two systems influence each other such that the state of one system systematically affects the dynamics of the other \cite{Ivanov2016}. Coupled systems are not operating independently; rather, the state of one system constrains the accessible states of the other. This concept extends beyond mere statistical correlation to imply mechanistic (causal) physiological interdependence.

Our findings meet the criteria for true physiological coupling through several lines of evidence:

\textbf{Statistical dependency and state constraint:}

Fetal HR complexity is not independent of mHR state. When mHR decelerates, fHR complexity systematically changes ($\beta$ = -0.123, p = 0.012), demonstrating that maternal physiological state constrains the fetal state space. Fetal HR dynamics become more predictable and regular during mHR decelerations, indicating fewer accessible dynamical states—a hallmark of coupling in complex systems \cite{Bashan2012}.

\textbf{Bidirectional asymmetric influence:}

We observe coupling in both directions—maternal states influence fetal dynamics (coupling strength 60\%) and fetal states influence maternal dynamics (coupling strength 40\%)—confirming true interdependence rather than unidirectional causation. The asymmetry in coupling strength (maternal$\to$fetal stronger than fetal$\to$maternal) indicates hierarchical physiological organization while preserving bidirectionality.

\textbf{Temporal coordination:}

The coupling effects occur during the physiological events (maternal/fetal HR decelerations), not at random times. Transfer entropy analyses reveal directional information flow with temporal lag, and our previous pPRSA work \cite{Lobmaier:2020} demonstrated phase-synchronized fetal responses to maternal events. This temporal specificity distinguishes coupling from coincidental co-occurrence.

With this conceptual foundation established, we can now examine the quantitative strength and mechanistic basis of this coupling.

\subsubsection{Quantifying Coupling Strength: The 60\% Ratio}

Our MLM analysis revealed that mHR decelerations exert substantially stronger influence on fHR complexity than any other physiological state. This finding arises from the mathematical relationship between the MLM beta coefficients, which may not be immediately apparent from examining the figures alone.

The conditioning framework MLM yields two critical coefficients:
\begin{enumerate}
    \item  Baseline (no conditioning): $\beta$ = +0.206, p < 0.001 — representing fetal entropy when measured independently of conditioning;
    \item  Maternal deceleration conditioning: $\beta$ = -0.123, p = 0.012 — representing entropy change during maternal bradycardic events.
\end{enumerate}

The ratio of these coefficients quantifies proportional coupling strength: 0.123 / 0.206 = 0.597 $\approx $60\%. This means the maternal deceleration coupling effect ($\beta$ = -0.123) captures 60\% of the dynamic range established by the no-conditioning baseline ($\beta$ = +0.206). This quantifies how strongly maternal bradycardic state constrains fHR complexity relative to the available entropy variation in our analysis.

This 60\% describes the ratio of beta coefficients (coupling strength), not a 60\% reduction in actual entropy rate values. The entropy rate decreases by 0.123 units during maternal decelerations; this magnitude represents 60\% of the 0.206-unit baseline coefficient, establishing maternal decelerations as the strongest coupling condition detected in our entire analysis.

The coupling strength ratio quantifies the effect relative to the baseline variation, providing a normalized measure of coupling intensity that facilitates comparison across different physiological states. This represents a fundamental coupling mechanism conserved across all pregnancies—both stressed and control groups exhibit this profound entropy reduction during maternal decelerations. This stress invariance is essential for interpreting the findings in the context of past studies.

The profound asymmetry in coupling strength—with maternal decelerations exerting substantially greater influence than maternal accelerations or fetal states—raises an important physiological question: Why should maternal HR decelerations be especially potent in constraining fetal HR dynamics?

\subsubsection{Universal mechanism vs. Stress-Modulated Response?}

One possibility relates to respiratory-cardiovascular coupling. Maternal HR decelerations often coincide with expiratory phases of the respiratory cycle, when vagal tone is maximal ~\cite{Grossman2007}. Our previous work ~\cite{Lobmaier:2020} linked maternal breathing patterns to fetal heart rate responses, suggesting that the coupling we observe may reflect respiratory-mediated autonomic fluctuations. During maternal expiration, the diaphragm position changes, potentially altering intra-abdominal and uterine pressure. Simultaneously, maternal vagal activation during expiration could affect uterine blood flow or transmission of autonomic signals.

A second possibility concerns oxygenation dynamics. Maternal bradycardia, even within normal ranges, may signal reduced cardiac output or altered gas exchange that have immediate relevance to fetal oxygenation. The fetus may have evolved heightened sensitivity to maternal bradycardic states as an early warning system for potential oxygenation challenges, resulting in the strong coupling we observe.

The asymmetry may also reflect autonomic regulatory priorities. Maternal decelerations represent increased parasympathetic activity, which in the mother associates with rest, recovery, and potentially reduced activity. Fetal coupling to these states might facilitate coordinated rest periods or represent an adaptation to changes in maternal physiological state that alter the intrauterine environment.

Our previous work using bivariate phase-rectified signal averaging (bPRSA) demonstrated that the Fetal Stress Index (FSI) was significantly higher in stressed pregnancies \cite{Lobmaier:2020}:

\begin{itemize}
    \item Stressed group: FSI = 0.43 (0.18-0.85), showing fetal HR decreases during maternal decelerations;
    \item Control group: FSI = 0.00 (-0.49-0.18), fetuses remained "stable" during maternal decelerations.
\end{itemize}

That is, a higher FSI implies a stronger fHR response, which indicates higher stress levels. This causes an apparent contradiction: How can the current study report a stress-invariant 60\% coupling strength reduction when \cite{Lobmaier:2020} reported a stress-sensitive FSI?

We posit that this contradiction is resolved when one considers the different aspects of coupling. FSI captures the magnitude of the fetal HR decrease and is stress-sensitive, while conditioned entropy captures fHR complexity reduction and is stress-invariant. Stress modulates fHR response amplitude which FSI captures, while the fHR complexity reduction we detected is a universal constraint mechanism. The entropy reduction (60\%) and the fHR response magnitude (FSI) capture complementary but distinct physiological phenomena:

\begin{enumerate}
    \item \textit{Entropy reduction} (current study - universal): Quantifies how predictable/constrained fetal HR becomes during maternal decelerations. This occurs in ALL pregnancies and represents a fundamental maternal-fetal coordination. Even if the fHR does not decrease dramatically (low FSI in controls), the signal complexity still drops—indicating that the fetus is "tracking" the maternal state.
    \item \textit{FSI magnitude} (\cite{Lobmaier:2020} - stress-sensitive): Quantifies how strongly the fHR actually decreases during maternal decelerations. This response magnitude is modulated by stress, with stressed fetuses showing larger decreases ("over-sensitization").

\end{enumerate}

Putting this together, we can say that the 60\% entropy reduction represents the "coupling channel"—a universal communication pathway between mother and fetus that exists regardless of stress. What stress modulates is the signal transmitted through this channel—specifically, the amplitude and directionality of fetal HR responses. The 60\% entropy reduction is analogous to the bandwidth of a communication channel (conserved architecture), while FSI measures the volume of signals transmitted (modulated by stress). What differs between healthy and stressed pregnancies is not this coupling capacity but rather the intensity and pattern of signals transmitted through this universal channel, as evidenced by stress-sensitive FSI and exploratory TE associations.

\subsection{Methodological Advances: The Three-Layer Conditioning Framework}

Our conditioning framework represents a conceptual advance in characterizing physiological coupling. Traditional bivariate coupling measures, including bPRSA, typically compare signals under different conditions or quantify overall coupling strength. The conditioning approach systematically dissects:

\begin{enumerate}
    \item Univariate properties: Baseline signal complexity independent of coupling
    \item Self-conditioned properties: How signals change during their own events
    \item Cross-conditioned properties: How signals change during the other signal's events (true bivariate coupling)
\end{enumerate}
This hierarchical approach is critical for interpretation. A change in fetal heart rate entropy during maternal decelerations could reflect: (a) the fetus happening to have its own decelerations simultaneously (self-conditioning), or (b) genuine constraint imposed by the maternal state (cross-conditioning). By quantifying both, we isolate true bivariate coupling from coincidental state matching.

The substantial coupling effect under cross-conditioning—with coupling strength of approximately 60\% during maternal decelerations—demonstrates that fHR complexity is genuinely constrained by maternal physiological states, not merely correlated due to independent fluctuations. This mechanistic specificity goes beyond correlation to reveal directional physiological influence.

\subsection{Clinical and Physiological Implications}

\textbf{Extending biomarker development:}

The Fetal Stress Index ~\cite{Lobmaier:2020} demonstrated clinical potential for identifying fetuses exposed to maternal stress. The information-theoretical framework developed here suggests that multiple complementary biomarkers may be needed. Transfer entropy may capture stress-sensitive temporal coupling, while conditioned entropy measures fundamental coordination mechanisms. A multimodal approach combining these measures could improve both sensitivity and mechanistic understanding.

\textbf{Neurodevelopmental prediction:}

While our exploratory entropy-neurodevelopment associations did not survive correction for multiple comparisons, the tentative patterns suggesting domain-specific relationships—transfer entropy with stress physiology, entropy rate with motor development, sample entropy with language—warrant investigation in larger samples. If replicated, epoch-specific entropy features computed during acceleration and deceleration events may prove more informative than full-recording features, consistent with the state-dependent nature of coupling we observed.

\textbf{Fundamental coupling mechanisms:}

The stress-invariant state-dependent coupling we identified may represent a universal maternal-fetal coordination mechanism, conserved across demographic variation and robust to maternal stress effects. This fundamental coupling could serve as a reference against which pathological conditions (severe stress, placental dysfunction, fetal growth restriction) could be evaluated. Deviations from normal state-dependent coupling might indicate compromised maternal-fetal communication requiring clinical attention.

\textbf{Sex-specific vulnerabilities:}

The exploratory observation of opposite coupling patterns in male and female fetuses—if replicated—could have implications for sex-specific clinical monitoring. Female fetuses might require different assessment approaches than male fetuses, with coupling measures potentially offering complementary information to traditional fetal monitoring parameters.

\subsection{Limitations and Future Directions}

\textbf{Sample size constraints:}

With 120 participants contributing 52 entropy features, our study was adequately powered for mixed linear model analyses with focused predictors but severely underpowered for multivariate modeling. The exploratory correlation analyses revealed numerous associations that did not survive correction for multiple comparisons. These null FDR results do not prove the absence of relationships—they reflect insufficient power for the large number of statistical tests performed. Future studies should employ sample sizes of n > 500 to enable robust multivariate analysis and adequately powered subgroup analyses while maintaining control for multiple comparisons.

\textbf{Neurodevelopmental follow-up:}

Bayley assessments were available for approximately 55\% of the cohort, with sample sizes for specific developmental domains ranging from 30-66 participants. While sufficient for exploratory correlation analysis, these samples were too small for definitive conclusions about entropy-development relationships. Complete follow-up in larger cohorts is essential for validating potential neurodevelopmental associations.

\textbf{Stress measurement:}

Our binary stress classification (stressed vs. control) based on maternal report may not capture the full stress continuum or distinguish chronic from acute stress effects. Future work should incorporate: continuous stress measures enabling dose-response analyses; physiological stress biomarkers beyond cortisol (inflammatory markers, autonomic function); longitudinal stress trajectories across pregnancy; and assessment of specific stress types (anxiety vs. depression vs. life events).

\textbf{Sample entropy limitations in brief windows:}

Our inability to detect bivariate coupling in sample entropy, despite similar effect directions to those of the entropy rate, highlights a methodological constraint. Sample entropy requires adequate data points for reliable estimation. Brief event windows (accelerations and decelerations lasting 2-10 seconds) contain insufficient data, resulting in algorithm failures and predominantly zero values. This limitation is not unique to our study but affects any complexity analysis of brief physiological events. Future research should explore: longer event windows by aggregating similar events; alternative complexity metrics less sensitive to sample-size constraints (multi-scale entropy, permutation entropy); higher sampling rates to increase data points per event; or hybrid approaches combining different entropy measures across different temporal scales.

\textbf{Mechanistic validation:}

While our information-theoretical analyses reveal coupling patterns, they do not directly identify the physiological mechanisms mediating maternal-fetal communication. Future work should combine entropy analyses with: simultaneous measurement of potential mediators (uterine blood flow, amniotic pressure, maternal oxygenation); experimental manipulations in animal models where mechanistic pathways can be directly tested; computational modeling to determine which physiological pathways can account for observed coupling patterns; and integration with placental function measures to assess the role of placental health in coupling strength.

\textbf{Replication imperative:}

The sex-stratified patterns and entropy-outcome associations reported here are exploratory, hypothesis-generating findings that did not survive correction for multiple comparisons. Independent replication in adequately powered samples with pre-registered analysis plans is essential before these patterns inform biological interpretation or clinical application. The robust findings—mixed linear model results for acceleration/deceleration asymmetry, conditioned entropy coupling, and sex-by-stress interactions in transfer entropy—provide the foundation for reliable biological inference and should be the focus of mechanistic follow-up.

\section{CONCLUSIONS}

Maternal-fetal coupling may operate through multiple physiological pathways (hemodynamic, hormonal, mechanical, neurohormonal) that create a communication channel between mother and fetus. Our findings demonstrate that:

\begin{enumerate}
    \item \textbf{Coupling is a mechanistic interdependence}, not mere correlation—maternal states constrain fetal dynamics through physiological pathways, meeting criteria for true coupled dynamical systems.
    \item \textbf{The coupling strength is asymmetric}, with maternal decelerations producing the strongest constraint on fetal HR complexity (60\% coupling strength), likely reflecting:
    \begin{itemize}
        \item Vagal dominance and respiratory coordination during maternal expiration
        \item Hemodynamic signals relevant to fetal oxygenation
        \item Coordinated rest states facilitating maternal-fetal synchronization
        \item Adaptive surveillance of potentially vulnerable maternal states
    \end{itemize}
    \item \textbf{The coupling mechanism is universal and stress-invariant} (60\% coupling strength in both stressed and control groups), representing conserved maternal-fetal communication architecture across normal pregnancy variation.
    \item \textbf{Stress modulates signals, not capacity}—the coupling channel (60\% coupling strength) is preserved, but stressed fetuses transmit stronger response signals (higher FSI) through this channel, consistent with autonomic over-sensitization.
    \item \textbf{Multiple measures capture complementary aspects}:
    \begin{itemize}
        \item Conditioned entropy: Coupling architecture/capacity (universal, stress-invariant)
        \item FSI/Transfer entropy: Signal transmission (stress-sensitive, variable)
    \end{itemize}
\end{enumerate}

These findings advance understanding of maternal-fetal physiological coupling, reveal potential biomarkers for stress and neurodevelopment, and demonstrate the value of multi-scale entropy analysis for characterizing complex physiological interactions.

\pagebreak


\appendix

\clearpage
\section*{Supplementary Tables}

\begin{table}[htb]
\caption{Complete MLM results for acceleration/deceleration analysis}
\label{tab:mlm_acc_dec_complete}
\begin{tabular}{l l l l l}\toprule
Effect & $\beta$ & SE & p-value & Sig \\\midrule

Event\_Type (deceleration) & -0.0606 & 0.0028 & <0.001 & *** \\
HR\_Source(maternal) $\times$ Event\_Type(decel) & +0.0282 & 0.0028 & <0.001 & *** \\
HR\_Source (maternal) & -0.0109 & 0.0028 & <0.001 & *** \\
Sex (male) & -0.0026 & 0.0028 & 0.361 & ns \\
Stress (stressed) & +0.0007 & 0.0027 & 0.802 & ns \\
Sex $\times$ Stress & -0.0001 & 0.0029 & 0.985 & ns \\
Sex $\times$ HR\_Source & +0.0027 & 0.0029 & 0.342 & ns \\
Sex $\times$ Event\_Type & +0.0014 & 0.0029 & 0.626 & ns \\
Stress $\times$ HR\_Source & -0.0005 & 0.0028 & 0.865 & ns \\
Stress $\times$ Event\_Type & -0.0013 & 0.0028 & 0.651 & ns \\ \bottomrule
\end{tabular}

\begin{flushleft}
\small\textit{Note.} Model: Fraction \~ Sex $\times$ Stress $\times$ HR\_Source $\times$ Event\_Type + (1|Patient\_ID)
REML estimation, n=120 patients, 480 observations
\end{flushleft}
\end{table}

\begin{table}[htb]
\caption{Complete MLM results for entropy rate conditioning analysis}
\label{tab:mlm_er_complete}
\begin{tabular}{l l l l l l}\toprule
Effect & $\beta$ & SE & p-value & Sig & Analysis Layer \\\midrule

Conditioning Effects &  &  &  &  &  \\
Conditioning (none) & +0.2061 & 0.0351 & <0.001 & *** & Univariate baseline \\
Conditioning (mother\_decel) & -0.1228 & 0.0491 & 0.012 & * & Cross-conditioned (bivariate) \\
Conditioning (fetus\_decel) & -0.0816 & 0.0424 & 0.054 & † & Cross-conditioned (bivariate) \\
Conditioning (mother\_accel) & -0.0335 & 0.0490 & 0.494 & ns & Cross-conditioned (bivariate) \\
Metric \& Signal &  &  &  &  &  \\
Metric (hmean) & -0.1172 & 0.0515 & 0.023 & * & - \\
HR\_Source (mother) & +0.0534 & 0.0337 & 0.113 & ns & - \\
Demographic &  &  &  &  &  \\
Sex (male) & -0.1058 & 0.0784 & 0.177 & ns & - \\
Stress (stressed) & -0.0852 & 0.0559 & 0.128 & ns & - \\
Sex $\times$ Stress & +0.1079 & 0.0886 & 0.223 & ns & - \\ \bottomrule
\end{tabular}
\begin{flushleft}
\small\textit{Note.} † p < 0.10 (marginal trend).
Model: Value \~ Sex $\times$ Stress $\times$ Metric $\times$ HR\_Source $\times$ Conditioning + (1|Patient\_ID).
REML estimation, n=120 patients, 1,262 observations
\end{flushleft}
\end{table}

\begin{table}[htb]
\caption{Entropy rate versus sample entropy: data quality comparison}
\label{tab:er_se_data_quality}
\begin{tabular}{l l l}
\toprule
Conditioning Type & Entropy Rate (Hmax/Hmean) & Sample Entropy \\
\midrule
Total observations & 1,262 & 286 (23\% of ER) \\
Obs per patient & 10.5 & 2.4 (23\% of ER) \\
fetus\_full & 100\% non-zero & 100\% non-zero \\
mother\_full & 100\% non-zero & 100\% non-zero \\
fetus\_fHR\_decel & \~100\% non-zero & 0\% non-zero  \\
mother\_fHR\_decel & \~100\% non-zero & 0\% non-zero  \\
fetus\_mHR\_accel & \~100\% non-zero & 13.3\% non-zero \\
mother\_mHR\_accel & \~100\% non-zero & 13.3\% non-zero \\
fetus\_mHR\_decel & \~100\% non-zero & 5.8\% non-zero \\
mother\_mHR\_decel & \~100\% non-zero & 5.8\% non-zero \\
\bottomrule
\end{tabular}
\end{table}

\begin{table}[htb]
\caption{Complete correlation matrix: Transfer Entropy features vs.\ cortisol}
\label{tab:te_complete}
\begin{tabular}{@{}lccc@{}}
\toprule
Feature & $r$ & $p$ & Significance \\
\midrule
Max TE fHR (all) & $+$0.083 & 0.437 & ns \\
Max TE fHR (accel) & $+$0.250 & 0.019 & * \\
Max TE fHR (decel) & $+$0.271 & 0.011 & * \\
Max TE mHR (all) & $+$0.083 & 0.437 & ns \\
Max TE mHR (accel) & $+$0.287 & 0.007 & ** \\
Max TE mHR (decel) & $+$0.315 & 0.003 & ** \\
Mean TE fHR (all) & $+$0.057 & 0.594 & ns \\
Mean TE fHR (accel) & $+$0.212 & 0.047 & * \\
Mean TE fHR (decel) & $+$0.207 & 0.053 & ns \\
Mean TE mHR (all) & $+$0.057 & 0.594 & ns \\
Mean TE mHR (accel) & $+$0.221 & 0.038 & * \\
Mean TE mHR (decel) & $+$0.217 & 0.042 & * \\
\bottomrule
\end{tabular}
\begin{flushleft}
\small\textit{Note.} ns = not significant; * $p < 0.05$; ** $p < 0.01$.
\end{flushleft}
\end{table}

\begin{table}[htb]
\caption{Transfer entropy MLM results: Conditioning and demographic effects}
\label{tab:te_mlm_complete}
\begin{tabular}{l l l l l}\toprule
Effect & $\beta$ & SE & p-value & Significance \\\midrule

Conditioning \& Metric &  &  &  &  \\
TE metric (Mean) & -0.0773 & 0.0073 & <0.001 & *** \\
HR event (None/Baseline) & -0.0374 & 0.0072 & <0.001 & *** \\
TE metric $\times$ HR event (None) & +0.0636 & 0.0102 & <0.001 & *** \\
Demographic Effects &  &  &  &  \\
Stress (stressed) & +0.0233 & 0.0105 & 0.026 & * \\
Sex $\times$ Stress & -0.0425 & 0.0164 & 0.009 & ** \\
Sex $\times$ Stress $\times$ HR event (None) & +0.0367 & 0.0149 & 0.014 & * \\
Non-Significant &  &  &  &  \\
Sex (male) & +0.0176 & 0.0111 & 0.113 & ns \\
Conditioning source (maternal) & +0.0050 & 0.0073 & 0.488 & ns \\ \bottomrule
\end{tabular}
\begin{flushleft}
\small\textit{Note.} Model specification: TE value $\sim$ Sex $\times$ Stress $\times$ TE\_type $\times$ Conditioning\_source $\times$ HR\_event + (1|Patient\_ID), REML estimation, n=120 patients.
\end{flushleft}
\end{table}

\clearpage
\section*{Supplementary Figures}

\begin{figure}[htb]
\begin{center}
\includegraphics[width=.8\linewidth]{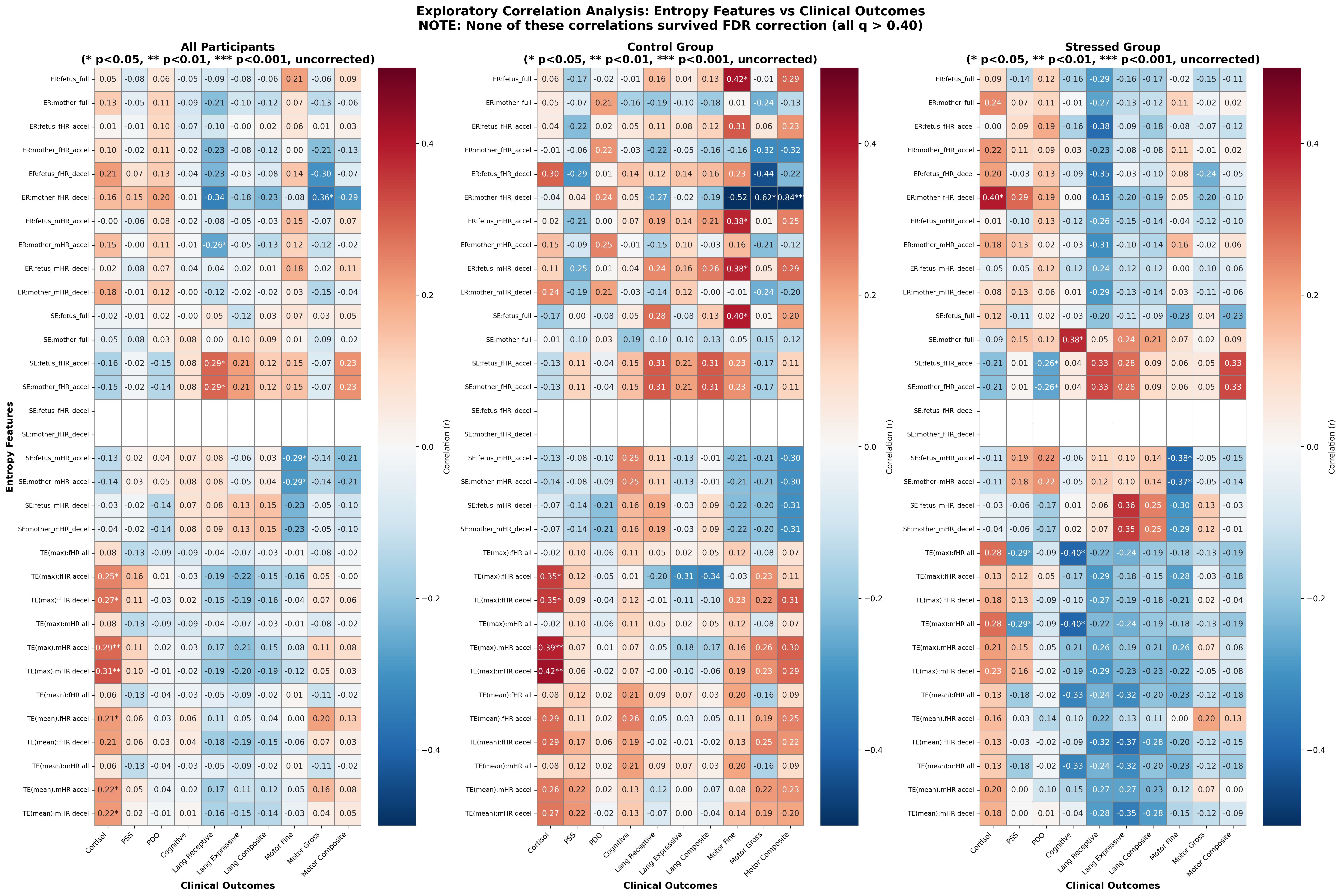}
\end{center}
\caption{Exploratory correlation analysis between entropy features and clinical outcomes.
}
\label{fig:heatmap_all}
\begin{flushleft}
\small\textit{Note.} Heatmaps show correlation coefficients (r) for all participants (left), control group (center), and stressed group (right). Asterisks indicate uncorrected significance (* p<0.05, ** p<0.01, *** p<0.001). \textbf{CRITICAL}: None of these correlations survived False Discovery Rate correction (all q > 0.40); all findings are exploratory and hypothesis-generating only.
\end{flushleft}
\end{figure}

\begin{figure}[htb]
    \centering
    \includegraphics[width=0.5\linewidth]{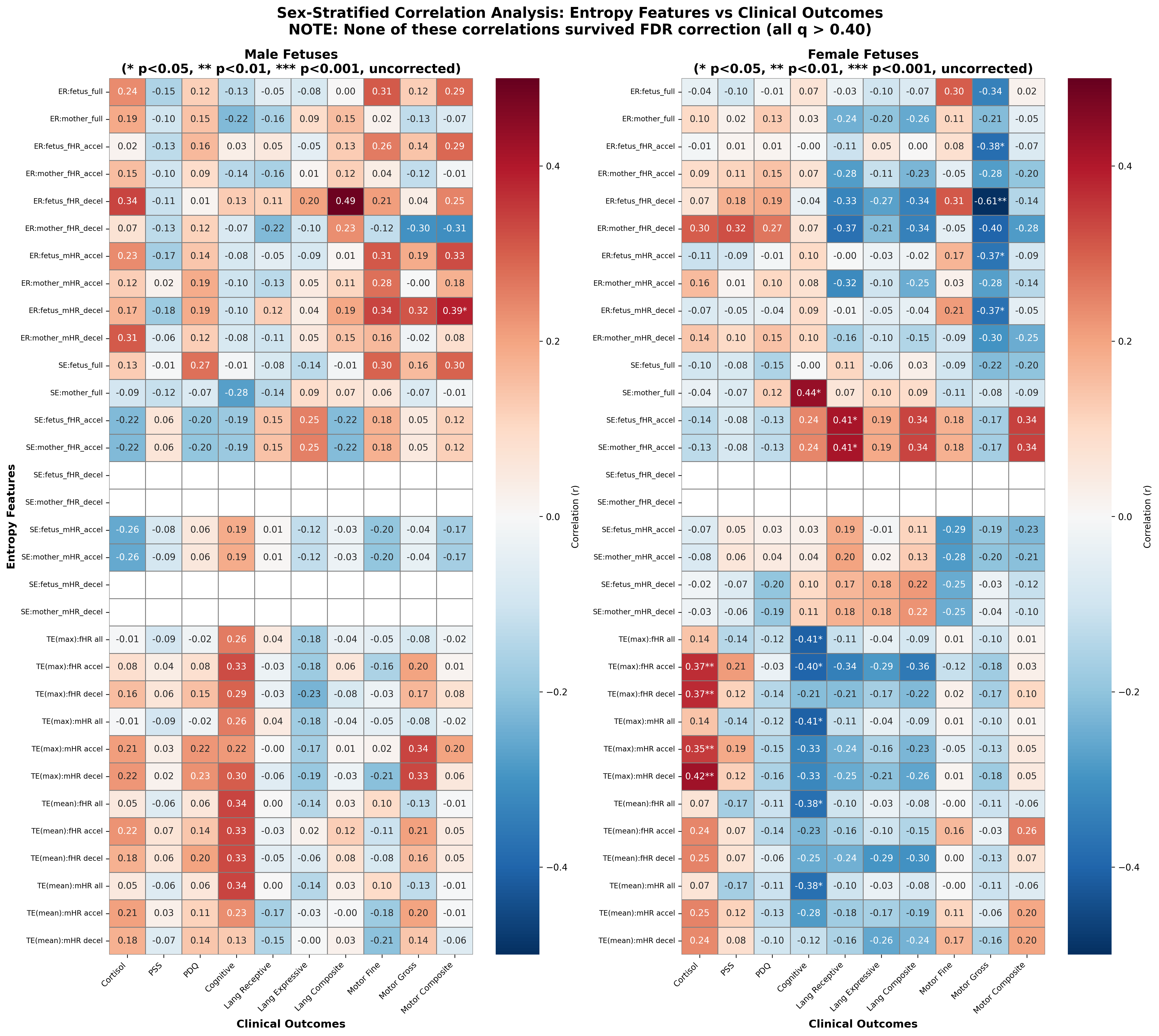}
    \caption{Sex-stratified sample entropy and entropy rate correlations to outcomes}
    \label{fig:se_er_sex_heatmap}
\begin{flushleft}
\small\textit{Note.} Heatmaps show correlation coefficients (r) for male fetuses (n=49, left), and female fetuses (n=71, right). Asterisks indicate uncorrected significance (* p<0.05, ** p<0.01, *** p<0.001). \textbf{CRITICAL}: None of these correlations survived False Discovery Rate correction (all q > 0.40); all findings are exploratory and hypothesis-generating only.
\end{flushleft}
\end{figure}

\begin{figure}[htb]
    \centering
    \includegraphics[width=0.5\linewidth]{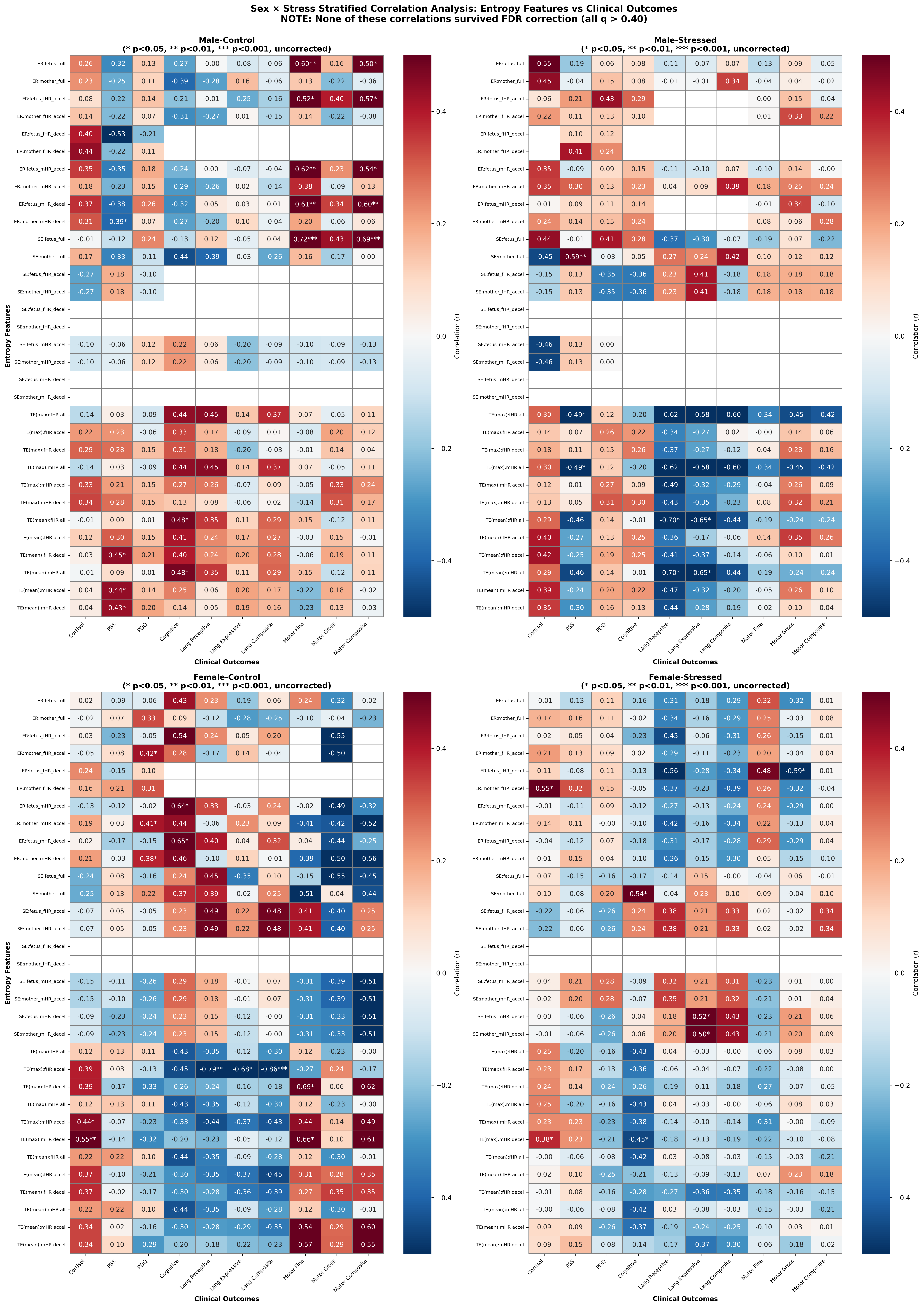}
    \caption{Sex $\times$ Stress stratified analysis of sample entropy and entropy rate correlations to outcomes}
    \label{fig:se_er_sex_stress_heatmap}
\begin{flushleft}
\small\textit{Note.} Exploratory sex $\times$ stress interaction analysis of entropy-outcome correlations. Four-panel heatmap (2$\times$2 grid) showing correlation coefficients (r) for: Male-Control (top-left, n=30), Male-Stressed (top-right, n=19), Female-Control (bottom-left, n=32), and Female-Stressed (bottom-right, n=39). Layout identical to Supplementary Figure \ref{fig:heatmap_all} with 32 entropy features (rows) $\times$ 10 clinical outcomes (columns). \textbf{CRITICAL}: None of these correlations survived False Discovery Rate correction (all q > 0.40); all findings are exploratory and hypothesis-generating only.
\end{flushleft}
\end{figure}

\clearpage
\section{SUPPLEMENTARY BOX: Understanding the Beta Coefficient Derivation and its Ratios as Coupling Strength}
\label{sec:coupling_box}

\textbf{For readers unfamiliar with MLM coefficient interpretation:}

Mixed linear models (MLM) with categorical predictors (like "Conditioning" with levels: none, mother\_accel, mother\_decel, fetus\_accel, fetus\_decel) use reference level coding. One category serves as the baseline (reference = 0), and beta coefficients represent differences from that reference.

In our MLM:

\begin{itemize}
    \item Reference level: fetus\_accel conditioning ($\beta$ = 0 by definition)
    \item No conditioning: $\beta$ = +0.206 (entropy is 0.206 units higher than fetus\_accel reference)
    \item Mother\_decel: $\beta$ = -0.123 (entropy is 0.123 units lower than fetus\_accel reference)
\end{itemize}
The 60\% calculation uses the no-conditioning coefficient as the baseline because it represents the maximal entropy state (independent measurement without event-specific conditioning). The maternal deceleration effect ($\beta$ = -0.123) is expressed as a proportion of this maximal entropy:

\textbf{Coupling strength} = $\displaystyle \frac{\mid \beta\,\, {\rm maternal\,\, decelerations}\mid}{\beta \,\, {\rm no\,\, conditioning}} = \frac{0.123}{0.206} = 0.597 \approx 60\%$

\medskip
This quantifies how much of the "available" entropy (i..e, dynamic range or the difference between the unconditioned state and the reference) is eliminated by maternal deceleration conditioning—revealing the magnitude of the constraint imposed by mHR decelerations on fHR dynamics. It's a measure of relative coupling strength, not absolute entropy percentage change.

Alternative calculation using predicted values:

\begin{itemize}
    \item Predicted entropy (no conditioning): 0.206 + reference\_value
    \item Predicted entropy (mother\_decel): -0.123 + reference\_value
    \item Difference: 0.206 - (-0.123) = 0.329
    \item Proportional reduction: 0.123 / 0.206 = 60\%
\end{itemize}
Both approaches yield the same conclusion: maternal decelerations reduce fetal entropy by 60\% relative to the unconditioned baseline, quantifying the profound asymmetric coupling in maternal-fetal HR dynamics. The 60\% coupling strength quantifies the effect relative to the baseline variation, not the absolute percentage change in entropy values.



\bibliographystyle{plain}
\bibliography{fetus}

\end{document}